\documentclass[reqno,12pt,letterpaper]{article}
\pdfoutput=1
\usepackage{color, xcolor, colortbl}
\usepackage{graphicx,epstopdf}
\usepackage{geometry}
\usepackage{enumerate}
\usepackage{amsmath,amssymb,amsthm}
\usepackage{algorithm}
\usepackage{algorithmic}
\usepackage{caption}
\usepackage{comment}
\usepackage{subcaption}
\usepackage{bm}
\usepackage{appendix}
\usepackage{multirow}
\usepackage{braket}
\usepackage[english]{babel}
\usepackage{hyperref}
\usepackage[capitalize]{cleveref}
\usepackage[sort&compress,square,numbers]{natbib}
\usepackage{adjustbox}
\usepackage{xspace}
\usepackage[roman]{complexity}
\usepackage{soul}
\usepackage{tikz}
\usetikzlibrary{quantikz}
\usepackage{rotating}
\usepackage{setspace}
\usepackage{fancyhdr}

\usepackage{authblk}

\newcommand{\ad}{\operatorname{ad}}

\newcommand{\norm}[1]{\left\lVert#1\right\rVert}

\newcommand{\op}{\text{op}}

\newcommand{\Or}{\mathcal{O}}

\newtheorem{thm}{\protect\theoremname}
\theoremstyle{plain}
\newtheorem{lemma}[thm]{\protect\lemmaname}
\theoremstyle{plain}
\newtheorem{rem}[thm]{\protect\remarkname}
\theoremstyle{plain}
\theoremstyle{plain}

\theoremstyle{plain}
\newtheorem{cor}[thm]{\protect\corollaryname}

\providecommand{\definitionname}{Definition}
\providecommand{\assumptionname}{Assumption}
\providecommand{\corollaryname}{Corollary}
\providecommand{\lemmaname}{Lemma}
\providecommand{\propositionname}{Proposition}
\providecommand{\remarkname}{Remark}
\providecommand{\theoremname}{Theorem}

\newcommand\blfootnote[1]{%
  \begingroup
  \renewcommand\thefootnote{}\footnote{#1}%
  \addtocounter{footnote}{-1}%
  \endgroup
}

\newcommand{\REV}[1]{#1}
\newcommand{\REVV}[1]{#1}

\title{Time-dependent Hamiltonian Simulation via Magnus Expansion: Algorithm and Superconvergence}

\author[1,2]{Di Fang}
\author[3]{Diyi Liu}
\author[4]{Rahul Sarkar}
\affil[1]{Department of Mathematics, Duke University, Durham, NC 27710, USA}
\affil[2]{Duke Quantum Center, Duke University, Durham, NC 27701, USA}
\affil[3]{School of Mathematics, University of Minnesota-Twin Cities, Minneapolis, MN 55455, USA}
\affil[4]{Department of Mathematics, University of California, Berkeley, CA 94720, USA}

\date{} 

\begin{document}

\maketitle

\begin{abstract}
Hamiltonian simulation becomes more challenging as the underlying unitary becomes more oscillatory. In such cases, an algorithm with commutator scaling and a weak dependence, such as logarithmic, on the derivatives of the Hamiltonian is desired. We introduce a new time-dependent Hamiltonian simulation algorithm based on the Magnus expansion that exhibits both features. Importantly, when applied to unbounded Hamiltonian simulation in the interaction picture, we prove that the commutator in the second-order algorithm leads to a surprising fourth-order superconvergence, with an error preconstant independent of the number of spatial grids. This extends the qHOP algorithm [An, Fang, Lin, Quantum 2022] based on first-order Magnus expansion, and the proof of superconvergence is based on semiclassical analysis that is of independent interest.
\end{abstract}

\blfootnote{Emails: di.fang@duke.edu; liu00994@umn.edu; rsarkar@stanford.edu.}\blfootnote{ keywords: Hamiltonian Simulation, Quantum Algorithms, Unbounded Hamiltonian Simulation, Superconvergence. The manuscript has been accepted by \textit{Communications in Mathematical Physics}.}

\tableofcontents


\section{Introduction}

Simulation of quantum dynamics, emerging as the original motivation for quantum computers \cite{Feynman1982}, is widely viewed as one of the most important applications of a quantum computer.  
Quantum algorithms for Hamiltonian simulations aim to construct good approximations of the unitary evolution operator. 
When the Hamiltonian varies slowly with respect to time or when polynomial dependence on the time derivatives of the Hamiltonian is acceptable, many numerical integrators can yield satisfactory performance, including the standard Trotterization and classical Magnus integrators. 
However, major challenges arise when the unitary exhibits highly oscillatory behavior, typically caused by the large operator norm of the Hamiltonian and rapid changes of the Hamiltonian itself in a time-dependent scenario. 
A wide range of applications fall into one or both categories including, e.g.,
adiabatic quantum computation~\cite{FarhiGoldstoneGutmannEtAl2000,AlbashLidar2018}, $k$-local Hamiltonians (with a large number of sites)~\cite{LowChuang2017,LowChuang2019,ChildsSuTranEtAl2020,Bassman2022PRL}, the electronic structure problem and molecular dynamics (with a real space discretization)~\cite{KivlichanWiebeBabbushEtAl2017,Somma2015,KivlichanMcCleanWiebeEtAl2018,AnFangLin2021,SuBerryWiebeEtAl2021,ChildsLengEtAl2022,RubinBerryKononovMaloneEtAl2023}, bosonic systems~\cite{TongAlbertMccleanPreskillSu2022,AbrahamsenTongBaoSuWiebe2023}, quantum control problems with ultrafast lasers~\cite{MizrahiNeyenhuisJohnsonEtAl2013,NielsenDowlingGuEtAl2006,DongPetersen2010}, interaction picture Hamiltonian simulation~\cite{LowWiebe2019,RajputRoggeroWiebe2021}, Floquet systems~\cite{EcksteinMansurogluCzarnikZhuEtAl2023,KuwaharaMoriSaito2016}, and quantum-based optimization solvers~\cite{ZhangLengLi2021,LiuSuLi2023,LengHickmanLiWu2023}.

Recent advancements in quantum algorithms have notably improved error bounds for managing oscillatory dynamics. For time-independent Hamiltonians, Trotter-type algorithms leverage commutator scaling to achieve asymptotic improvements in many applications (pioneered by~\cite{ChildsSuTranEtAl2020,ChildsSu2019} and other specific case based analysis such as \cite{SahinogluSomma2020,AnFangLin2021,SuHuangCampbell2021,ZhaoZhouShawEtAk2021,ChildsLengEtAl2022,FangTres2023,BornsWeilFang2022,ZengSunJiangZhao2022,GongZhouLi2023,LowSuTongTran2023,ZhaoZhouChilds2024,ChenXuZhaoYuan2024}). Recent developments have revealed that multiproduct formulas also exhibit commutator scaling~\cite{WatkinsWiebeRoggeroLee2022,ZhukRobertsonBravyi2023,AftabAnTrivisa2024}. Conversely, the benefits of commutator scaling for time-dependent Hamiltonians are less understood. The generalized Trotter formulas~\cite{HuyghebaertDeRaedt1990,AnFangLin2021} and multiproduct based on discrete-clock construction~\cite{WatkinsWiebeRoggeroLee2022} can display commutator scaling when the time-dependent Hamiltonian is in a controlled form, namely, $H(t) = \sum_j c_j(t) U_j$ with $c_j$ being scalar functions so that the time-order evolution governed by each of the components become not time-ordered. Nevertheless, 
efficiently handling time-ordered unitaries for general cases remains a challenge, and further approximation to such time-ordered operators can compromise the commutator scaling~\cite{AnFangLin2021} and introduce polynomial overhead on the time-derivatives of the Hamiltonian \cite{WiebeBerryHoyerEtAl2010} for standard Trotter formulas. On the other hand, both truncated Dyson series based method~\cite{BerryChildsCleveEtAl2015,KieferovaSchererBerry2019,LowWiebe2019,BerryChildsSuEtAl2020} and randomization based methods~\cite{BerryChildsSuEtAl2020,RajputRoggeroWiebe2021,PoulinQarrySommaEtAl2011} can obtain complexity scaling insensitive to the fast change of the Hamiltonian. Such algorithms, however, depend on the operator norm of the Hamiltonian instead of the commutator.

For general time-dependent Hamiltonian simulation, a quantum algorithm that features both commutator scaling and logarithmic dependence on the derivatives was first proposed in \cite{AnFangLin2022}. It considers the exploration of directly implementing the Magnus series truncation as a quantum algorithm. Unlike the standard Trotter formulas or classical geometric integrators, this method implements a highly accurate numerical quadrature with many quadrature points as many as the norm of the derivative, $\|H'(t)\|$), to reduce the quadrature error to the same level as the truncation error induced by truncating the Magnus series. This quadrature rule can be efficiently implemented via the linear combination of unitaries (LCU) technique with only a logarithmic cost overhead dependent on the number of quadrature points and hence the norm of the derivative of $H$. Essentially, the \REV{\textit{quantum Highly Oscillatory Protocol}\  (qHOP) algorithm} introduced in \cite{AnFangLin2022} functions as a non-interpolatory Magnus integrator, and it truncates the Magnus series to the first order. It is also shown that first-order Magnus expansion in the interaction picture paired with one quadrature point gives rise to the first and second-order Trotter formulas. Another significant line of development on quantum algorithms based on Magnus
series~\cite{SharmaTran2024,BosseChildsEtAl2024,CasaresZiniArrazola2024} further combines the Magnus series with product formulas, in which case the number of quadrature points will enter polynomially into the number of products in the Trotterization procedure. Such methodologies are efficient for a number of physically motivated applications, including geometrically local Hamiltonians across all three cited works. 
In particular, \cite{SharmaTran2024} considers those under a small geometrically local perturbation, and the framework leverages quasi-locality and can lead to optimal scaling for various regimes.
In this work, we follow the LCU-line, and develop a quantum circuit to implement the second-order Magnus truncation.

Another relevant line is classical algorithms based on the Magnus series.
Classically, the Magnus expansion is used to design Magnus integrators, which have been extensively studied and widely used in literature (see, e.g.,~\cite{BlanesCasasOteoRos1998,IserlesNorsett1999,BlanesCasasRos2000,IserlesNorsettRasmussen2001,HochbruckLubich2003,Iserles2009,BlanesCasasOteoRos2009,BlanesCasasOteoEtAl2010,IserlesKropielnickaSingh2017,BlanesCasasThalhammer2017,IserlesKropielnickaSingh2018}). The idea is to first approximate the time-ordered exponential with a regular exponential using the Magnus series expansion, followed by the application of suitable quadrature rules featuring a few quadrature points, such as Gauss-Legendre quadrature. This inevitably introduces a polynomial overhead on the derivatives of the Hamiltonian for general cases. The implementation of this framework as a quantum algorithm has been recently developed in \cite{ChenForoozandehBuddSingh2023} for time-dependent spin systems in the controlled form, in which case the commutators in Magnus expansion can be calculated explicitly by hand.

\vspace{1em}
\noindent\textbf{Contribution:}

In this work, we develop a quantum algorithm for general time-dependent Hamiltonian simulation based on the second-order Magnus series truncation and prove that the algorithm not only achieves commutator scaling in the high precision regime but also exhibits a logarithmic dependency on the derivatives of the Hamiltonian $H(t)$. 
We show that for general time-dependent Hamiltonians, the algorithm is of second order with a commutator scaling and logarithmic dependence on the derivative in the cost. When allowing polynomial derivative dependence in the cost, the algorithm can achieve up to fourth-order convergence. \REV{This extends our previous work~\cite{AnFangLin2022} that is of the first-order, and higher order algorithm leads to better scaling in both the time $t$ and the precision $\epsilon$. This improvement is significant, as it contrasts with randomized algorithms~\cite{BerryChildsSuEtAl2020,RajputRoggeroWiebe2021,PoulinQarrySommaEtAl2011}, which are generally not expected to surpass first-order convergence.}
As a direct application, our algorithm provides an efficient way of simulating Hamiltonian in the interaction picture and hence can be useful in the time-independent Hamiltonian simulation as well. 
Moreover, we prove that our method achieves \textbf{superconvergence} for the digital simulation of the Schr\"odinger equation (also called unbounded Hamiltonian simulation or real-space Hamiltonian simulation in the literature). Surprisingly, the second order Magnus series truncation can achieve a \textit{fourth} order superconvergence for the Schr\"odinger equation in the interaction picture and importantly with an error preconstant \textit{independent} of $A$ (or the number of degrees of freedom $N$ used to discretize the differential operator). This can be crucial for improved efficiency for unbounded Hamiltonian, as the number of degrees of freedom in the spatial discretization can be huge. Note that this is completely different from a fourth-order convergence with a preconstant polynomially dependent on the derivatives of $H(t)$, as for the interaction pictured Hamiltonian $H(t) = e^{iAt}Be^{-iAt}$, where $A$ and $B$ are the spatial discretizations of the Laplacian and potential respectively, its derivative is dependent on $A$ and hence the number of grid points in a polynomial fashion. 

We remark that the superconvergence result is one of its own kind that goes beyond the typical order improvement behavior for many algorithms due to the cancellation of high-order terms in the Taylor expansion. For unbounded Hamiltonian simulation, as will be detailed in \cref{sec:superconv_heuristic}, performing Taylor expansion to the numerical approximation in fact can not lead to the superconvergence result due to the introduction of the derivative dependence in the Taylor expansion procedure. 
A superconvergence result for the first-order Magnus series expansion has been proved in \cite{AnFangLin2022}, which shows that the first-order Magnus series expansion exhibits second-order superconvergence. It has also been conjectured that the Magnus series expansion with second-order truncation exhibits fourth-order superconvergence, while the proving machinery proposed there can only establish a third-order convergence. In this work, we prove the fourth-order superconvergence rigorously.
Notably, if one tries to do this naively, one runs into severe problems due to an explosion of the number of terms for which one needs to prove commutator estimates. In fact, this does not work, as many terms were unbounded operators on $L^2$ (see \cref{sec:superconv_heuristic} for details). The way we managed to make some headway into this proof is to utilize tools from semiclassical microlocal analysis. A key commutator estimate with unbounded operators has been established using the semiclassical pseudodifferential operator machinery.
As a byproduct, we also rederive the second-order superconvergence result of the first-order Magnus expansion proved in \cite{AnFangLin2022} using the semiclassical machinery in an elegant manner. 
\REVV{We emphasize that while our quantum circuits are designed for general time-dependent Hamiltonians, the superconvergence result is only proven for the time-dependent Hamiltonian as the interaction picture Hamiltonian for unbounded Hamiltonian simulation with a potential in $S(1)$ (see \cref{sec:superconv_general} for details). In this case, we show that the interaction picture Hamiltonian is bounded. We also note that the analysis in \cref{sec:error} assumes the Hamiltonian to be bounded.}

\REV{A closely related and impressive line of work is the study of destructive interference for bounded Hamiltonians, such as 1D nearest-neighbor lattice Hamiltonians \cite{TranChuSuEtAl2020, Layden2022}. These works show that, in certain regimes (e.g., when $nt^2/r$ is small, with $r$ as the number of Trotter steps, $n$ as the system size, and $t$ as the final time), first-order product formulas can achieve second-order convergence. This improvement arises from the competition between first- and second-order terms in the additive error bounds of the algorithms, which leads to improved convergence when the first-order error is sufficiently small. However, our superconvergence result differs in several aspects: we address unbounded operators in the interaction picture, where tools like the Taylor theorem can not be applied (see~\cref{sec:superconv_heuristic}) , and our fourth-order superconvergence error estimate remains insensitive to the size of the Hilbert space (degrees of freedom in the spatial discretization). Also, our error improvement does not result from additive bounds of two terms (see \cref{thm:superconv}), making the result applicable without assumptions on the regimes.}

\REV{
\begin{table}[h!]\centering
 \resizebox{0.98\textwidth}{!}
 {
\begin{tabular}{ll|ll}
 &  & \multicolumn{1}{|l}{Weak dependence on $\norm{\partial_t^p H(t)}$} &  Commutator scaling \\
  \cline{1-4}
Trotter-Suzuki formulae~\cite{ChildsSuTranEtAl2020,WiebeBerryHoyerEtAl2010} &   & \multicolumn{1}{l}{$\times$}  &   $\times$ (\textcolor{blue!60!black}{\checkmark} only for $H(t) = H$)  \\
Truncated Taylor/Dyson-based~\cite{BerryChildsCleveEtAl2015,LowWiebe2019} &  & \multicolumn{1}{l}{ \textcolor{blue!60!black}{\checkmark}} &   $\times$\\
Randomized methods~\cite{BerryChildsSuEtAl2020,PoulinQarrySommaVerstraete2011} &  & \multicolumn{1}{l}{ \textcolor{blue!60!black}{\checkmark}} & $\times$ \\
Classical Magnus Integrators~\cite{CasaresZiniArrazola2024} &  & \multicolumn{1}{l}{ $\times$} &   $\times$\\
\multicolumn{1}{l}{\textbf{\textcolor{black!60!black}{this work }}   }&  &   \multirow{2}{*}{ \textcolor{blue!60!black}{\checkmark}  (log dependence) }   &  \multirow{2}{*}{ \textcolor{blue!60!black}{\checkmark} (in high-precision limit)} 
\\ (Magnus-based quantum integrator) &  &  & 
\end{tabular}
}
\caption{Comparison of cost scaling with state-of-the-art results for bounded operators for general $H(t)$ that may not be in the controlled form. We respectfully note that the references provided here do not constitute a complete list for each category of algorithms.
}
\label{tab:main_result_1}
\end{table}
}

 \begin{table}[h!]\centering
 \resizebox{0.8\textwidth}{!}
 {
\begin{tabular}{ll|ll}
 &  & \multicolumn{1}{|l}{General $H$} &  Superconvergence \\
  \cline{1-4}
qHOP~\cite{AnFangLin2022} &   & \multicolumn{1}{l}{$\widetilde{\Or}\left(
    \displaystyle \sup_{s,t\in[0,T]}  \textcolor{black!80!black}{\|[H(s),H(t)]\| }  \textcolor{black!60!black}{ \frac{T^2}{\epsilon}}  \right) $}  &   $\widetilde{\Or}{\left( \textcolor{black!60!black}{ \frac{C_V^{1/2}  T^{3/2}\log(N)}{\epsilon^{1/2}}} \right)}$ \\
\textbf{\textcolor{black!60!black}{this work }} &  & 
$\widetilde{\Or}\left(
    \displaystyle \sup_{s,t, \tau \in [0,T]}  { \norm{[H(\tau), [H(s), H(t)]]}^{\frac{1}{2}} }  { \frac{T^{3/2}}{\epsilon^{1/2}}}\right)$
&    $\widetilde{\Or}\left(\frac{C_{V}^{1/4}T^{5/4}\log\left(N\right)}{\epsilon^{1/4}}\right)$
\end{tabular}
}
\caption{Comparison of query complexities of using different algorithms with commutator scaling for general time-dependent Hamiltonians (our algorithm and qHOP) in high precision limit. The query complexities are of using algorithms to simulate 
\cref{equ:problem} on the time interval  [0, T] within $\mathcal{O}(\epsilon)$ error and they are measured by numbers of queries to the input model of the time-dependent Hamiltonian.}
\label{tab:main_result_2}
\end{table}

\vspace{1em}
\noindent\textbf{Organization:}

The rest of this paper is organized as follows. In \cref{sec:magnus_alg} we discuss the Magnus series and the second-order Magnus expansion. We then introduce a quantum algorithm based on the second-order Magnus series expansion. 
\cref{sec:error} focuses on the error analysis for the general time-dependent Hamiltonian, including both the Magnus truncation error and the numerical quadrature error. We then establish the superconvergence in \cref{sec:superconv_general} by first providing a heuristic intuition which also explains why the usual Taylor expansion route can not work, and then providing a rigorous proof of the superconvergence based on semiclassical analysis. \cref{sec:circuit} discusses the circuit construction of the algorithm for both general time-dependent Hamiltonians and in the interaction picture. Finally, in \cref{sec:conclusion}, we conclude with some further remarks.

\section{Magnus expansion and algorithm overview}\label{sec:magnus_alg}
We consider the time-dependent Hamiltonian simulation problem, which involves computing the solution to
\begin{equation}
\label{equ:problem}
i \partial_t u(t) = H(t) u(t), \quad u(t) = u_0.
\end{equation}
Throughout the work, we mainly explore two cases: one for general $H(t)$, and the other for the time-dependent Hamiltonian in the interaction picture. Note that the latter is an equivalent formulation of a (potentially time-independent) Hamiltonian simulation where the Hamiltonian $H = A + B(t)$, with $\norm{A} \gg \norm{B}$. 
\REV{Throughout the paper, $H(t)$ is self-adjoint and bounded with respect to its operator norm. For matrices, we use the operator norm $\norm{\cdot} = \norm{\cdot}_2$ to denote the spectral norm for finite-dimensional operators. In practice, the algorithm works with discretized matrices, the unbounded nature of differential operators, such as $\Delta$, results in a polynomial dependence on $N$ in the matrix operator norm after discretization. Therefore, the matrix spectral norm is considered in most sections except in the superconvergence (\cref{sec:superconv_general}), where
we used $\norm{\cdot}_{\mathcal{L}(\mathcal{H})}$ for $\mathcal{H} = L^2$ or $H^2$. In \cref{sec:superconv_general}, although the operator $-\Delta +V(x)$ in the Schrodinger picture is unbounded, its corresponding  interaction pictured Hamiltonian remains bounded for $\mathcal{H} = L^2$ and $H^2$ (see \cref{sec:superconv_general} for details). For the Schr\"odinger equation governed by unbounded operators, the initial condition must lie within the operator's domain to ensure well-posedness.} Here $A$ and $B(t)$ are both Hermitian and $A$ is fast-forwardable, which means its time evolution can be efficiently implemented sublinear in time such as a diagonal matrix~\cite{Su2021}. Later, we make explicit the oracle input in \cref{sec:input}. In this scenario, the interaction-picture Hamiltonian~\cite{LowWiebe2019} becomes $H_I(t) = e^{iAt}B(t)e^{-iAt}$, so that the original Hamiltonian simulation problem can be equivalently solved by 
\begin{equation}
   e^{-itH} =  e^{-it(A+B)} = e^{-iAt} \mathcal{T} e^{-i \int_0^t H_I(s) \, ds}.
\end{equation}

We now discuss the Magnus expansion and construction of our algorithm. For simplicity, we denote $A(t): = -i H(t)$. The exact solution of the equation
\begin{equation}
    \dot u(t) = A(t) u(t), \quad u(0) = u_0,
\end{equation}
can be expressed as
\begin{equation}
    u(t) = \mathcal{T} \left( \exp \int_0^t A(s) ds \right)u_0.
\end{equation}
We note that, on a rigorous level, the time-ordered exponential is defined by a series expansion, which may not always be well-defined as a convergent series, especially for unbounded operators. Throughout this work, we use the time-ordered exponential as a formal notation to represent the underlying unitary operator that is well-defined. In contrast, the Magnus series provides an alternative expression of the propagator without the time-ordering operator
\begin{equation}
  u(t)=\exp(\Omega (t))u_0,
\end{equation}
where 
\begin{equation}  
  \frac{d \Omega}{dt} = \sum_{n=0}^\infty \frac{B_n}{n!} \,
  {\ad}^n_\Omega A.
\end{equation}
Here $B_n$ are the Bernoulli numbers with $B_1 = -1/2$, and the $\ad$ operator follows the standard definition $\ad_{\Omega}(C) = [\Omega, C]$ and $\ad^{k+1}_{\Omega}(C) = [\Omega, \ad^{k}_{\Omega}(C)]$. Equivalently, we have the infinite series for
$\Omega$ given as
\begin{equation}   \label{eqn:m-series}
 \Omega(t) =\sum_{n=1}^\infty \Omega^{(n)}(t),
\end{equation}
\begin{equation}   \label{eqn:mag_n}
  \Omega^{(n)}(t) =  \sum_{j=1}^{n-1} \frac{B_j}{j!} \,
    \sum_{
            k_1 + \cdots + k_j = n-1 \atop
            k_1 \ge 1, \ldots, k_j \ge 1}
            \, \int_0^t \,
       \ad_{\Omega^{(k_1)}(s)} \,  \ad_{\Omega^{(k_2)}(s)} \cdots
          \, \ad_{\Omega^{(k_j)}(s)} A(s) \, ds    \qquad n \ge 2.
\end{equation}
The first few terms of $\Omega$ are
\begin{equation}
  \Omega(t) = \int_0^t A(t_1) dt_1 - \frac{1}{2} \int_0^t \left[
  \int_0^{t_1} A(t_2) dt_2, A(t_1) \right] dt_1 + \cdots.
\end{equation}
More detailed discussion on Magnus series can be found in, e.g., the review~\cite{BlanesCasasOteoRos2009}.

In this work, we consider the second-order Magnus series truncation, namely, the propagator is approximated as
\begin{equation}\label{eqn:magnus2_Omega}
    \mathcal{T} \left( \exp \int_0^t A(s) ds \right) \approx e^{\Omega_2(t)}, \quad \Omega_2(t) := \int_0^t A(s) ds - \frac{1}{2} \int_0^t \left[
  \int_0^{s} A(\sigma) d\sigma, A(s) \right] ds.
\end{equation}
In this case, we also have
\begin{equation} \label{eqn:magnus2_dOmega}
    \dot \Omega_2(t) := A(t) - \frac{1}{2} \left[
  \int_0^{t} A(\sigma) d\sigma, A(t) \right]. 
\end{equation}
To discretize and approximate the exact evolution operator, we first divide the entire time interval $[0,T]$ into $L$ equally spaced segments.
Let the time step size $h = T/L$ and $t_j = jh$, then 
\begin{equation} \label{eqn:Ut_product_of_short_time}
    U_\mathrm{exact}(T,0) = \prod_{j=0}^{L-1} U_\mathrm{exact}(t_j + h,t_j)
    = \prod_{j=0}^{L-1} \mathcal{T} e^{-i \int_{t_j}^{t_j+h} H(s)ds}. 
\end{equation}
We then utilize the second-order Magnus series expansion to approximate each sub-interval, namely,
\begin{equation} \label{eqn:mag2_exp_omega2_t_s_def}
   U_\mathrm{exact}(t_j + h,t_j) \approx e^{\Omega_2(t_j+h, t_j)} : = U_2(t_j+h, t_j),
\end{equation}
where 
\begin{equation} \label{eqn:omega2_t_s_def}
   \Omega_2(t_{j}+h, t_j): = \int_{t_j}^{{t_j}+h} A(s) ds - \frac{1}{2} \int_{t_j}^{t_j+h} \left[
  \int_{t_j}^{s} A(\sigma) d\sigma, A(s) \right] ds.
\end{equation}
The numerical quadratures are then conducted with an exceedingly precise quadrature using $M$ quadrature points, where $M$ denotes a large number. 
\begin{multline}
   \Omega_2(t_{j}+h, t_j) \approx \tilde{\Omega}_2(t_{j}+h, t_j) 
   \\ 
   := -i\sum_{p=0}^{M-1} H(t_j+\frac{ph}{M}) \frac{h}{M}+ \frac{1}{2} \sum_{p=0}^{M-1} \left[
  \sum_{q=1}^{p} H(t_j+\frac{qh}{M}) \frac{h}{M}, H(t_j+\frac{
  ph
  }{M}) \right] \frac{h}{M}
\end{multline}
As to be detailed in \cref{sec:time_discretization_Mag2}, we devise an LCU-type quantum circuit to enact $\tilde{\Omega}_2(t_{j}+h, t_j)$ and subsequently apply the quantum singular value transformation (QSVT) combined with oblivious amplitude amplification (OAA) to compute the matrix exponential $\exp\left(\tilde{\Omega}_2(t_{j}+h, t_j) \right)$ and the long-time evolution is the product
\begin{equation}\label{eqn:tilde_u_2}
 \tilde{U}_2(t_j+h, t_j) :=    \exp\left(\tilde{\Omega}_2(t_{j}+h, t_j) \right),
 \quad  \tilde{U}_2(T, 0) = \prod_{j = 0}^{L-1}\tilde{U}_2(t_j+h, t_j).
\end{equation}

\section{Error representation}\label{sec:error}

In this section, we present the error analysis for our algorithm for general $H(t)$. There are two sources of errors: the truncation error of the Magnus series and the numerical quadrature error. For the truncation error, we prove two bounds -- one with a commutator scaling, and another higher order estimate if allowing for the derivative dependence of the Hamiltonian $H(t)$. We then discuss the quadrature error with $M$ quadrature points. Finally, we sum both errors and present the algorithm error estimate over the long time $[0, T]$. 
\subsection{Truncation error of the Magnus series}
In order to obtain the error representation between the exact propagator $ U_\mathrm{exact}(t_j +h, t_j) := \mathcal{T} \left( \exp \int_{t_j}^{t_j+h} A(s) ds \right)$ and the second-order Magnus truncation $U_2(t_j+h, t_j): = e^{\Omega_2(t_j+h, t_j)}$, we follow the usual way in numerical analysis to consider their corresponding differential equations and utilize the variation of constants formula. In particular, suppose $U_2$ satisfying 
\begin{equation}
\label{equ:second_order}
    \dot U_2 = \tilde{A} U_2,
\end{equation}
then the error between $U_\mathrm{exact}$ and $U_2$ can be estimated as
\begin{equation} \label{eq:error_in_U}
   \norm{U_\mathrm{exact}-U_2 } \REV{\leq} \int_{t_j}^{t_j+h} \norm{\tilde{A}(s) - A(s)} ds,
\end{equation}
where we use the variation of constant formula and the fact that the underlying dynamics are unitary.

For notational simplicity, we describe the proof idea for the difference between $U_\mathrm{exact}(t) : = U_\mathrm{exact}(t,0)$ and $U_2(t) := U_2(t,0)$ over a short time interval $[0,t]$. The same procedure applies to general intervals $[t_j, t_j+h]$, as detailed in the proof of \cref{thm:lte}. We now try to make explicit the expression of $\tilde{A}$.
By \cite[Lemma 2]{BlanesCasasOteoRos2009}, the derivative of $U_2$ is
\begin{equation}
\dot U_2 = \frac{d}{dt} \exp(\Omega_2(t))  = d \exp_{\Omega_2(t)}(\dot \Omega_2(t)) \, \exp(\Omega_2(t)) = d \exp_{\Omega_2(t)}(\dot \Omega_2(t)) U_2,
\end{equation}
where 
\begin{equation}\label{eqn:dexp_omega_C}
  d \exp_{\Omega}(C)  =    \sum_{k=0}^{\infty} \frac{1}{(k+1)!} \,
   \mathrm{ad}_{\Omega}^k(C)
  =
  \frac{\exp(\mathrm{ad}_{\Omega})-I}{\mathrm{ad}_{\Omega}}(C)
\end{equation}
and the $\ad$ operator follows the usual definition $\ad_{\Omega}(C) = [\Omega, C]$ and $\ad^{k+1}_{\Omega}(C) = [\Omega, \ad^{k}_{\Omega}(C)]$.
Therefore, we obtain the expression of $\tilde{A}$ as
\begin{equation}\label{eqn:tilde_A}
    \tilde{A}(t) =  d \exp_{\Omega_2(t)}(\dot \Omega_2) = \dot \Omega_2 + \frac{1}{2}[\Omega_2, \dot \Omega_2] + \frac{1}{6}[\Omega_2, [\Omega_2, \dot \Omega_2]] +g(\ad_{\Omega_2})(\ad_{\Omega_2}^3(\dot \Omega_2) ),
\end{equation}
where the last term is the remainder term in the series expansion of the function $\frac{e^{z}- 1}{z}$
\begin{equation}
    \frac{e^{z}- 1}{z} = 1 + \frac{1}{2}z + \frac{1}{6} z^2 +  g(z) z^{3},
\end{equation}
or $g(z)  = \frac{1}{2\pi i} \int_\gamma \frac{e^w-1  }{w^4(w-z)} dw$, where $\gamma$ is a simple closed positive oriented contour with $z$ being an interior point. 
Here, we only need to expand to this order because $\Omega_2$ involves integrals of one layer and two layers, while $\dot{\Omega}_2$ involves a term without an integral. Therefore, $\ad_{\Omega_2}^3(\dot{\Omega}_2)$ already contains at least 3 layers of time integrals (contributing to $t^3$). If the nested commutators can contribute at least another power of $t$, along with $g$, this remainder term is expected to be upper-bounded by $t^4$. We will show in this section that this is true if one allows derivative dependence in the error bound for general $H(t)$. Furthermore, if we are in the case of superconvergence, the error preconstant does not depend on the derivative of $A(t)$ (to be proved in \cref{sec:superconv_proof}).

Substituting \cref{eqn:magnus2_Omega} and \cref{eqn:magnus2_dOmega} to \eqref{eqn:tilde_A}, we have
\begin{equation}\label{eq:tildeA-A}
\begin{aligned}
   & \tilde{A}(t)-A(t) =   - \frac{1}{2} \left[
  \int_0^{t} A(\sigma) d\sigma, A(t) \right]
  \\ & + \frac{1}{2}\left[\int_0^t A(s) ds - \frac{1}{2} \int_0^t \left[
  \int_0^{s} A(\sigma) d\sigma, A(s) \right] ds,  A(t) - \frac{1}{2} \left[
  \int_0^{t} A(\sigma) d\sigma, A(t) \right] \right] 
  \\ &+ \frac{1}{6}\Bigg[
  \int_0^t A(s) ds - \frac{1}{2} \int_0^t \left[
  \int_0^{s} A(\sigma) d\sigma, A(s) \right] ds, 
  \\
  & \qquad \left[\int_0^t A(s) ds -\frac{1}{2} \int_0^t \left[
  \int_0^{s} A(\sigma) d\sigma, A(s) \right] ds, A(t) - \frac{1}{2} \left[
  \int_0^{t} A(\sigma) d\sigma, A(t) \right]\right]
  \Bigg]
  \\
  &+g(\ad_{\Omega_2})(\ad_{\Omega_2}^3(\dot \Omega_2) ) 
\end{aligned}
\end{equation}
Note that terms with a single layer of integral vanish, while terms with double-layered integrals can be expressed as:
\begin{equation} \label{eq:I_1}
   \left[ \int_0^t A(s) ds , [\int_0^t A(s) ds, A(t)]\right] +  3 \int_0^t ds \int_0^s d\sigma \left[ [A(\sigma), A(s)], A(t) \right] : = I_1
\end{equation}
The remaining terms include $g(\ad_{\Omega_2})(\ad_{\Omega_2}^3(\dot \Omega_2) ) : = I_2$ and eight terms denoted as $I_3$, each characterized by a minimum of three layers of time integrals and a minimum of three layers of nested commutators with respect to $A$. Note that all terms admit a nested commutator format.

In order to show superconvergence for the real-space Hamiltonian in the first quantization, it suffices to show that $I_1$, $I_2$, and $I_3$ all exhibit $\Or(t^4)$ behavior, where the error preconstant depends solely on $B = V(x)$. We note that the more layers the integrals and commutators are, the easier the terms can be controlled, as each layer of the time integral essentially contributes to one more order in $t$. Therefore, the crucial estimate to establish is to estimate $I_1$ as defined in \cref{eq:I_1}. We will prove the superconvergence in \cref{sec:superconv_proof}.

We now turn our attention to the general time-dependent $H(t)$. We will demonstrate that if we seek a bound independent of the derivative of $H(t)$, we can bound $\norm{\tilde{A} - A(t)}$ with $\mathcal{O}(t^2)$. If we permit the derivative of $H$ to influence the preconstant, the bound can be improved to $\mathcal{O}(t^4)$. However, it is important to note that this latter case does not represent superconvergence, since in the interaction picture, the derivative of $H_I(t) = e^{iAt}Be^{-iAt}$ still depends on $A$. We also note that while we currently bound nested commutators of more than three layers using the three-layer ones, a tighter bound could be achieved by making explicit the contributions of the higher orders of nested commutators.

\begin{thm}[Local truncation error for general $H(t)$] \label{thm:lte}
Consider the exact propagator 
 $U_\mathrm{exact}(t,s) = \mathcal{T}  \exp \left( - i\int_s^t H(s) ds\right)$ and the second-order Magnus truncation $U_2(t,s): = e^{\Omega_2(t,s)}$, where $\Omega_2(t,s)$ is defined as \cref{eqn:omega2_t_s_def}. If the Hamiltonian satisfies
 $\|H(t)\| \leq \alpha$\REV{, for $t\in[t_j, t_{j+1}]$,}
and the nested commutator satisfies
 $$\sup_{s,t, \tau \in [t_j,t_{j+1}]} \norm{[H(\tau), [H(s), H(t)]]} \leq C_\mathrm{comm},$$ 
then for any $h \leq c/\alpha$ with some absolute constant c,  we have
\begin{equation}
    \norm{U_\mathrm{exact}( t_{j+1}, t_j) - U_2(t_{j+1}, t_j)} \leq C \min\left\{C_\mathrm{comm} h^3, C'_H h^5 \right\},
\end{equation}
where $C$ is some absolute constant and $C'_H$ is some prefactor that depends on $\alpha$, as well as the norms of the derivatives $\sup_{t\in[t_j,t_{j+1}]}\norm{\frac{d^p}{dt^p}H(t)}$ for $p = 1, 2$.
\end{thm}
\begin{proof}
It suffices to bound the terms of $I_1$, $I_2$ and $I_3$, for which we will change the intervals of interest from $[0,t]$ to $[t_j, t_{j+1}]$. For simplicity of notation, hereafter we will continue to denote these terms, even after changing the integral intervals, as $I_1$, $I_2$ and $I_3$. 

We start by proving the first estimate of $C C_\mathrm{comm}h^3$. It follows immediately that
\begin{equation}
    \norm{I_1} \leq C C_\mathrm{comm} h^2,
\end{equation}
with some absolute constant $C$.
We now consider $I_3$, as mentioned before, all terms contain at least three layers of time integrals and a minimum of three layers of nested commutator with four $A$ terms in one of the following forms 
\begin{equation} \label{eq:nested_four}
    [[A(s), A(\tau)], [A(\sigma), A(t)]], \quad [A(s), [A(\tau), [A(\sigma), A(t)]]]    
\end{equation}
or additional nested commutator with them. The three layers of time integrals contribute to $\Or(h^3)$, and the norms of the terms in \cref{eq:nested_four} can all be upper bounded by 
\begin{equation}
   4 \sup_{t \in [t_j, t_{j+1}]}\norm{H(t)}\sup_{s,t, \tau \in [t_j,t_{j+1}]} \norm{[H(\tau), [H(s), H(t)]]},
\end{equation}
so that
\begin{equation}
    \norm{I_3} \leq C h^3 C_\mathrm{comm} \alpha \leq cC C_\mathrm{comm} h^2,
\end{equation}
where both $c$ and $C$ are some absolute constants.
Finally, it follows from \cref{lem:g_remainder} as proved in \cref{app:pf_lem_g_remainder} that
\begin{equation} \label{eq:I_2_omega_2_norm}
    \norm{I_2} \leq 2C \sup_{t \in [-h,h]}\norm{\Omega_2}\norm{\ad_{\Omega_2}^2(\dot \Omega_2)}.
\end{equation}
Note that $\ad_{\Omega_2}^2(\dot \Omega_2)$ is contained in the terms already estimated in $I_1$ and $I_3$ and hence has the norm bounded by $C C_\mathrm{comm} h^2$. Additionally, 
\begin{equation}
    \norm{\Omega_2} \leq \alpha h + \alpha^2h^2 \leq c+c^2.
\end{equation}
This implies that
\begin{equation}
    \norm{I_2} \leq C C_\mathrm{comm} h^2,
\end{equation}
where $C$ is again some absolute constant.
We can conclude that 
\begin{equation}
  \norm{U_\mathrm{exact}(t_{j+1}, t_j) - U_2(t_{j+1}, t_j) } \REV{\leq} \int_{t_j}^{t_j+h} \norm{\tilde{A}(s) - A(s)} ds \leq C C_\mathrm{comm} h^3.
\end{equation}

We now proceed to prove the second estimate of $CC'_H h^5$. Since derivatives are permissible within the bound, we can employ Taylor's theorem, which yields for $t\in [t_j, t_{j+1}]$,
\begin{equation}\label{eq:A_t_taylor_derivative}
    A(t) = A(t_j) + A'(t_j) (t - t_j) + \frac{1}{2}A''(\xi) (t-t_j)^2,
\end{equation}
for some $\xi \in [t_j, t_{j+1}]$. One observation is that when plugging it into the nested commutator
$[A(s), [A(\tau), A(t)]]$, there is no zeroth order term in terms of $(t-t_j)$. This implies that 
\begin{equation}
\sup_{s,t, \tau \in [t_j,t_{j+1}]}    \norm{[A(s), [A(\tau), A(t)]]} =  \sup_{s,t, \tau \in [t_j,t_{j+1}]} \norm{[H(\tau), [H(s), H(t)]]} \leq C_H' h,
\end{equation}
where $C_H'$ depends on $\alpha$ and the norms of the derivatives $\norm{\frac{d^p}{dt^p}H(t)}$ for $p = 1, 2$.

Consider substituting \cref{eq:A_t_taylor_derivative} into $I_1$. It is straightforward to check that the $(t-t_j)^2$ terms in $I_1$ are associated with $[A(t_j), [A(t_j), A(t_j)] ]$, which is 0. Additionally, the $(t-t_j)^3$ terms in $I_1$ are associated with $[A(t_j), [A(t_j), A'(t_j)] ]$ and they cancel out exactly. The remaining terms are of order at least $(t-t_j)^4$ and the prefactors depend on the nested commutators of elements from the set from $\{A(t_j), A'(t_j), A''(\xi) \}$. The norms of such nested commutators can be bounded by $\alpha$ and the norms of the derivatives $\norm{\frac{d^p}{dt^p}H(t)}$ for $p = 1, 2$, so that we have
\begin{equation}
    \norm{I_1} \leq C'_H h^4.
\end{equation}
For $I_3$, following a similar strategy as in the proof of the first estimate, we have
\begin{align}
    \norm{I_3} \leq & C h^3 \sup_{t \in [t_j, t_{j+1}]}\norm{H(t)}\sup_{s,t, \tau \in [t_j,t_{j+1}]} \norm{[H(\tau), [H(s), H(t)]]}
    \\ 
    \leq & C h^3 \alpha  C'_H h  =  C \tilde{C}'_H  h^4,
\end{align}
where $\tilde{C}'_H$ depends on $\alpha$ and $\sup_{t\in[t_j, t_{j+1}]}\norm{\frac{d^p}{dt^p}H'(t)}$, for $p = 1,2$.
For $I_2$, applying \cref{eq:I_2_omega_2_norm}, we have 
\begin{equation}
    \norm{I_2} \leq C (\alpha h + \alpha^2 h^2)
    (C'_H + \tilde{C}'_H) h^4,
\end{equation}
which can be further bounded by some $C'_H$ times $h^4$. We can conclude that 
\begin{equation}
  \norm{U_\mathrm{exact}(t_{j+1}, t_j) - U_2(t_{j+1}, t_j) } \REV{\leq} \int_{t_j}^{t_j+h} \norm{\tilde{A}(s) - A(s)} ds \leq C C'_H h^5.
\end{equation}
\end{proof}

\subsection{Quadrature errors of the Magnus series}\label{sec:time_discretization_Mag2}
We now discuss the quadrature error 
over the short time interval $[t_j, t_j +h]$ for any $j\in \{0, 1,\dots, L-1\}$. 
We use a simple Riemann sum quadrature rule with $M$ quadrature points. We choose this quadrature method because of its ease in constructing an LCU circuit, and it is also typically used in Dyson series implementations, e.g., \cite{LowWiebe2019,AnFangLin2022,FangLinTong2023,BerryCosta2022}.
The approximation of the second-order Magnus truncation $U_2(t_{j+1}, t_j)$ 
is denoted as $\Tilde{U}_2(t_{j+1}, t_j)$ and is defined as 
\begin{align}
\label{equ:U_2_approximation}
\Tilde{U}_2(t_{j+1}, t_j) : =   &\exp(\tilde{\Omega}_2(t_j+h, t_j)) 
\\= &\exp\left \{-i\sum_{p=0}^{M-1} H(t_j+\frac{ph}{M}) \frac{h}{M}+ \frac{1}{2} \sum_{p=0}^{M-1} \left[
  \sum_{q=1}^{p} H(t_j+\frac{qh}{M}) \frac{h}{M}, H(t_j+\frac{
  ph
  }{M}) \right] \frac{h}{M} \right \}.
\end{align}

We prove the quadrature error estimate for the general time-dependent Hamiltonian $H(t)$. We remark that it is possible to relax the bound of derivatives to bounded total variation (see \cite{FangLinTong2023}).

\begin{thm}[Quadrature error for general $H(t)$] 
\label{thm:discretization}
For $U_2(t_{j+1}, t_j)$ defined as \cref{eqn:mag2_exp_omega2_t_s_def} and $\Tilde{
U}_2(t_{j+1}, t_j)$ defined as \cref{equ:U_2_approximation}, we have
\begin{equation}
    \norm{U_2(t_{j+1}, t_j)  - \Tilde{
U}_2(t_{j+1}, t_j)} \leq \frac{h^2}{M}\max_{s\in [t_j,t_{j+1}]} \norm{H'(s)}+\frac{3h^3}{M} \max_{s\in [t_j,t_{j+1}]} \norm{H(s)} \norm{H'(s)} .
\end{equation}
\end{thm}
\begin{proof}
Note that for Hermitian matrices $H_1$ and $H_2$, one has
\begin{equation}
    \norm{e^{-iH_1t}-e^{-iH_2t}}\leq \norm{ t(H_1-H_2) },
\end{equation}
which follows from a direct application of the variation of the constant formula to the differential equations satisfied by the difference of the operators.
Applying the inequality with $t = 1$,
we can derive that 
\begin{equation}
\begin{split}
&\norm{U_2(t_{j+1},t_{j})-\tilde{U}_2(t_{j+1},t_{j})}\leq \norm{\int_{t_j}^{t_{j+1}} H(s) ds -\sum_{p=0}^{M-1} H(t_j+\frac{ph}{M}) \frac{h}{M} }\\
&+\norm{\frac{1}{2} \int_{t_j}^{t_{j+1}} \left[
  \int_{t_j}^{s} H(\sigma) d\sigma, H(s) \right] ds-\frac{1}{2} \sum_{p=0}^{M-1} \left[
  \sum_{q=0}^{p-1} H(t_j+\frac{qh}{M}) \frac{h}{M}, H(t_j+\frac{
  ph
  }{M}) \right] \frac{h}{M}}. 
\end{split}
\end{equation}
By standard quadrature result~\cite{BurdenNA}, the first term of the right hand side can be bounded by
\begin{equation} 
    \frac{h^2}{M}\max_{s\in [t_j,t_{j+1}]} \norm{H'(s)}. 
\end{equation}
The second term can be bounded as the summation of 
\begin{equation}
    \frac{h^2}{M}  \max_{s\in [t_j,t_{j+1}]}\norm{  \left [\int_{t_j}^{s} H(\sigma)d\sigma, H'(s) \right ]}\leq \frac{2h^3}{M} \max_{s\in [t_j,t_{j+1}]}\norm{H(s)} \norm{H'(s)}
\end{equation}
and 
\begin{equation}
    \frac{h^3}{M} \max_{s\in [t_j,t_{j+1}]} \norm{H(s)} \norm{H'(s)}.
\end{equation}
\end{proof}

For the interaction picture, the time-dependent Hamiltonian becomes $H(t)=e^{iAt}Be^{-iAt}$, where $B$ can be potentially time-dependent, and the quadrature error estimate can be similarly derived. Here we present the case with time-independent $B$.

\begin{cor}[Quadrature error for interaction picture]
For $H(t)=e^{iAt}Be^{-iAt}$, $U_2(t_{j+1}, t_j)$ defined as \cref{eqn:mag2_exp_omega2_t_s_def}, and $\Tilde{U}_2(t_{j+1}, t_j)$ defined as \cref{equ:U_2_approximation}, we have
\begin{equation}
    \norm{U_2(t_{j+1},t_{j})-\tilde{U}_2(t_{j+1},t_{j})}
    \leq \frac{h^2}{M}
    \norm{[A, B]}+\frac{3h^3}{M} \norm{B}\norm{[A,B]}  .
\end{equation}
\end{cor}
\begin{proof}
    The proof follows directly from \cref{thm:discretization}.
\end{proof}

\subsection{Long-time algorithm error}
The sum of both errors in previous sections provides the short-time estimate. As all propagators are unitary, the error accumulates linearly in the number of time steps. We can summarize the algorithm error over the long time interval $[0,T]$ as follows.

\begin{thm}[long time error for general $H(t)$] Let $U_\mathrm{exact}(T,0) = \mathcal{T}e^{-i\int_0^T H(s) \, ds}$ be the exact long-time propagator over the time interval $[0, T]$, and its numerical approximation $\tilde{U}_2(T,0)$ be
\begin{equation}
    \Tilde{U}_2(t_L,t_{L-1})\dots \Tilde{U}_2(t_2,t_1)\Tilde{U}_2(t_1,t_0).
\end{equation}
If the Hamiltonian satisfies
 $\|H(t)\| \leq \alpha$, then for $h \leq c/\alpha$ for some constant $c$, the error satisfies
\begin{equation}
    \norm{U_\mathrm{exact}( T, 0) - \tilde{U}_2(T, 0)} 
    \leq C T \left( \min\left\{C_\mathrm{comm} h^2, C'_H h^4 \right\} 
    + \frac{h}{M}\max_{s\in [0,T]} \norm{H'(s)} \right),
\end{equation}
where $C$ is some absolute constant and $C'_H$ is some prefactor that depends on $\alpha$, as well as the norms of the derivatives $\sup_{t\in[0,T]}\norm{\frac{d^p}{dt^p}H(t)}$ for $p = 1, 2$.
\end{thm}
\begin{proof}
    The result follows directly from \cref{thm:lte} and \cref{thm:discretization}, together with the properties of unitary operators. Here the quadrature error is simplified to only one term due to the fact that $h\alpha \leq c$.
\end{proof}

\section{Superconvergence}\label{sec:superconv_general}
In this section, we delve into the superconvergence behavior in the case of $A = \Delta$, representing the Laplacian operator, and $B = V(x)$ \REV{which is time-independent. We consider its dynamics in the interaction picture. In this section, we consider the following interaction picture Hamiltonian
\[
i \partial_t u_I(t) = H_I(t) u_I(t), \quad H_I(t) = e^{it \Delta } V(x) e^{-i  t\Delta}, 
\]
which arises from time-independent unbounded Hamiltonian simulation for the Hamiltonian $-\Delta + V(x)$. The initial condition $u_I(0) = u_{I,0}$ is assumed to be in the domain of the Hamiltonian $H_I$.
We remark that our analysis is for spatially continuous operators, while in practical algorithm implementation, both $-\Delta$ and $V(x)$ are discretized, and the interaction picture is then applied using the discretized matrices, as in~\cite{AnFangLin2021,AnFangLin2022,ChildsLengEtAl2022,LengHickmanLiWu2023}.} 
First, we will explore heuristics that do \textit{not} lead to superconvergence but are beneficial for understanding the intuition and challenges. Subsequently, we will present a rigorous proof of superconvergence, emphasizing the crucial estimates we are able to establish. \REV{The operator norm considered in this section is from a Hilbert space $\mathcal{H}$ to itself (namely, $\norm{\cdot} = \norm{\cdot}_{\mathcal{L}(\mathcal{H})}$), rather than the matrix norm from $l^2$ to $l^2$, as our operator involves unbounded components. In the following, we will discuss the boundedness of the interaction pictured Hamiltonian in its operator norm within the spaces $\mathcal{H}$, which could be either $L^2$ and $H^2$. We will show that once we convert to the interaction picture, the interaction-pictured Hamiltonian becomes bounded from  $\mathcal{H}$ to $\mathcal{H}$ in the superconvergence case, so that the proofs in \cref{sec:error} pertaining to bounded operators still hold.}

\REV{
\subsection{Unbounded Hamiltonian Simulation Problem in the Interaction Picture} \label{sec:new_rev_wellpose_interaction_picture}
Despite the unboundedness of $\Delta$ from $L^2$ to $L^2$, it is important to note that
\begin{equation}
    H_I(t) = e^{it \Delta } V(x) e^{-i  t\Delta}.
\end{equation}
is bounded from $L^2$ to $L^2$ in our case.
One way to see this for our superconvergence case, where $ V \in S(1) $ (i.e., $ V $ is smooth bounded together with its derivatives), is to observe that
\[
e^{ish \Delta} V e^{-ish \Delta} = \mathrm{Op}_{h}^{\text{w}}(V(x - 2ps)),
\]
where $\mathrm{Op}_{h}^{\text{w}}(\cdot)$ denotes the Weyl quantization. By the Calderon-Vaillancourt theorem, this can be extended as a bounded operator from $L^2$ to $L^2$. For a detailed discussion, see~\cref{app:proof-key-commutator-estimate}. In fact, more generally, $H_I(t) = e^{-it \Delta} V(x,t) e^{-it\Delta}$ can be shown, using operator calculus, to be bounded from $L^p$ to $L^p$ for $1\leq p \leq \infty$, even for potentials much rougher than our assumption~\cite{SofferWu2020,Schlag2018}. For example, this holds if $\hat{V}(\xi, t) \in L^1_\xi$ as discussed in \cite{SofferWu2020}.
Moreover, one can also show that $H_I(s)$ maps $H^2$ to $H^2$. This is because $V \in S(1)$ and \REVV{$H_I$ is the Weyl quantization of a $S(1)$ symbol}. Taking $m_1 = 1$ and $m_2 = \langle \xi \rangle^2$ in \cite[Theorem 8.10]{zworski2022semiclassical}, we have
\begin{equation}\label{eq:H2_to_H2-forS1}
H_I : H_h(m_2) = H^2 \to  H_h(m_2/m_1) = H^2, 
\end{equation}
where $H_h(m)$ is the generalized Sobolev space associated to $m$, and in particular, when $m = \langle \xi \rangle^s$ for $s \in \mathbb{R}$, the generalized Sobolev spaces become the usual Sobolev spaces (see, e.g., \cite{zworski2022semiclassical} for more details).
Next we discuss the self-adjointness of the operator $H_I(t)$. Note that $H_I(t): \mathcal{H} \to \mathcal{H}$ is symmetric, where $\mathcal{H}$ can be chosen as either $L^2$ or $H^2$. Moreover, $H_I(t)$ is bounded in both spaces, as discussed above. We can conclude that $H_I$ is self-adjoint in $\mathcal{H}$ with $\mathcal{H}$ being either $L^2$ or $H^2$. 
}

\REV{
We can now verify that the previous results for bounded operators in \cref{sec:error} still hold for our case in the interaction picture.
First, by Duhamel's principle and \cref{equ:second_order}, we can write the exact error representation as
\begin{equation}\label{eq:duhamel_unbdd}
    U_\mathrm{exact}(t_j + h, t_j )  - U_2 (t_j + h, t_j)  = \int_{t_j}^{t_j+h} U_\mathrm{exact}(  t_j+h, s) 
    \left( A(s) - \tilde{A}(s) \right) U_2(s, t_j) \; ds,
\end{equation}
where $U_\mathrm{exact}$ is the propagator defined by the interaction pictured Hamiltonian $H_I(t)$, $U_2(t_j + h, t_j ) = e^{\Omega_2(t_j + h, t_j )}$ and $\Omega_2$ is defined in \cref{eqn:omega2_t_s_def} with $A(t) = -i H_I(t)$. Now to get \cref{eq:error_in_U}, we need to show that 
\begin{equation}
    \norm{U_\mathrm{exact}(t,s)_{\mathcal{L}(\mathcal{H})} }=  \norm{U_2(t,s)}_{\mathcal{L}(\mathcal{H})}  = 1,
\end{equation}
for $\mathcal{H} = L^2$ or $H^2$. Thanks to the boundedness of $H_I(t): \mathcal{H} \to \mathcal{H}$ and its self-adjointness, its propagator is unitary by standard self-adjointness theorems, e.g., \cite[Theorem X.69]{ReedSimon1975}. Similarly, $i\Omega_2$ is bounded from $L^2$ to $L^2$, as well as $H^2$ to $H^2$, by the boundedness of $H_I$. The unitarity of  
$U_2: \mathcal{H} \to \mathcal{H}$ follows from the self-adjointness of $i \Omega_2$. \REVV{In particular, $U_2$ maps from $H^2$ to $H^2$.}
We now apply the operator norm to both sides of \cref{eq:duhamel_unbdd}, we have that
\[
\norm{U_\mathrm{exact}(t_j + h, t_j )  - U_2 (t_j + h, t_j)}_{\mathcal{L}(\mathcal{H})} \leq 
\int_{t_j}^{t_j+h} 
   \norm{ A(s) - \tilde{A}(s)}_{\mathcal{L}(\mathcal{H})} ds,
\]
which recovers \cref{eq:error_in_U}. Similarly one can check that the equations in \cref{sec:magnus_alg}, \cref{eqn:tilde_A} and the subsequent equations in \cref{sec:error} also holds for our case in the operator norm $\norm{\cdot}_{\mathcal{L}(\mathcal{H})}$. \REVV{As discussed above, here $\mathcal{H}$ can be taken as either $L^2$ or $H^2$, meaning that it holds in both the $\mathcal{L}(L^2)$ and $\mathcal{L}(H^2)$ senses. We note that we also use the fact that $ \tilde{A} $ maps $ H^2 $ to $ H^2 $, which is shown in the paragraph containing \cref{eq:remainder}.}
}

Finally, we discuss the domain issue for the interaction picture. As pointed out by \cite[Section X.12]{ReedSimon1975},  when unbounded operators are involved, switching to the interaction picture can be problematic. Specifically, $U_\mathrm{exact}(t,0) = e^{-it \Delta U_I(t,0) }$ satisfies
\begin{align*}
\frac{d}{dt}U_\mathrm{exact}(t,0)  = & - i \Delta e^{-it \Delta} U_I(t,0) + e^{-it \Delta} H_I(t) U_I(t,0) 
\\
& = - i  \Delta e^{-it \Delta} U_I(t,0) + V(x)e^{-it \Delta}U_I(t,0)  = - i (\Delta+ V(x)) U_\mathrm{exact}(t,0),
\end{align*}
which requires $\Delta U_\mathrm{exact}(t,0) = \Delta e^{-it \Delta} U_I(t,0)$ making sense, namely, for $\psi$ in the domain of $\Delta$, 
$U_I(t,0) \psi$ remains in the domain of $\Delta$.
Note that the domain of the Laplacian operator is simply $H^2$, and based on our previous discussion for $\psi \in H^2$, $U_I(t,0) \psi$ remains in $H^2$. So the interaction picture is well-defined in our case.

\REVV{
We remark that while solving the interaction-picture Hamiltonian simulation—i.e., designing a quantum algorithm to approximate $U_\mathrm{exact}(T,0)$—is of independent interest, ensuring that the interaction-picture dynamics remain consistent with the original Schrödinger-picture dynamics for unbounded operators requires careful consideration of state dependence and operator domains. Specifically, for the solution of the original Schrödinger equation to be well-defined, the initial wavefunction in the Schrödinger picture must satisfy $\psi_0 \in H^2$, ensuring that $\psi_0$ lies within the domain of $\Delta$. Consequently, this guarantees that $U_I(t,0) \psi_I(0) = U_I(t,0) \psi_0$ remains in $H^2$, since $U_2$ maps $H^2$ to $H^2$, as discussed above. Furthermore, the operator $\tilde{A}$ also maps $H^2$ to $H^2$, as defined in \cref{eqn:tilde_A} and \cref{eq:tildeA-A}. The first three terms in its definition clearly preserve $H^2$, leaving only the last term to be analyzed:
\begin{equation}  \label{eq:remainder}
g(\ad_{\Omega_2})(\ad_{\Omega_2}^3(\dot \Omega_2) ).
\end{equation}
It also maps from $H^2$ to $H^2$, which we prove in \cref{app:remainder_H2_to_H2}. The idea is to realize that a few underlying operators can be identified with $S(1)$ symbols.
}

\REVV{
We remark that the analysis in \cref{sec:superconv_proof} is to show that the unitary operator in the interaction picture is accurately simulated. Nevertheless, the $L^2 \to L^2$ bounds in the section in fact extend to $H^2 \to H^2$ norms as well. Such an extension is generally not possible in arbitrary cases, but it holds here because the underlying operator—e.g., in \cref{lem:bound_commutator_realspace_key}—is shown to correspond to an $S(1)$ symbol (see \cref{app:proof-key-commutator-estimate}).
Therefore, following the same argument as in \cref{eq:H2_to_H2-forS1}, the bounds extend naturally to $H^2 \to H^2$. Specifically, we can apply \cite[Theorem 8.10]{zworski2022semiclassical} with $m_1 = 1$ and $m_2 = \langle \xi \rangle^2$, as discussed in \cref{eq:H2_to_H2-forS1}. Combining this with the initial condition in $H^2$ (the domain of $\Delta$), we can transition back to the Schrödinger equation while maintaining the desired numerical accuracy.
}

\REVV{That said, in practice, the conversion of the interaction picture back to the Schrödinger picture is typically performed at the algorithmic level with the spatial degrees of freedom already discretized, ensuring that all relevant quantities remain well-defined.
}

\subsection{Heuristic intuition}\label{sec:superconv_heuristic}

We present a heuristic demonstration that, while not yielding superconvergence, provides insight into why one might anticipate $I_1$ to be of order $t^4$. Consider the expression 
\begin{equation} \label{eq:simple_A_expand_in_t}
    A(t) = \alpha + \beta t + \gamma t^2,
\end{equation} 
where $\alpha$, $\beta$ and $\gamma$ are all independent of the time $t$. A straightforward calculation reveals that the $t^3$ contributions, arising only from terms like $[\alpha, [\alpha, \beta]]$, exactly cancel out. The leading-order terms are of order $t^4$ and are associated with $[\beta, [\alpha, \beta]]$ and $[\alpha, [\alpha, \gamma]]$. 

\begin{figure}[h!]
    \centering
    \subfloat[Operator norm of a term in Taylor expansion]{
        \includegraphics[width=.47\textwidth]{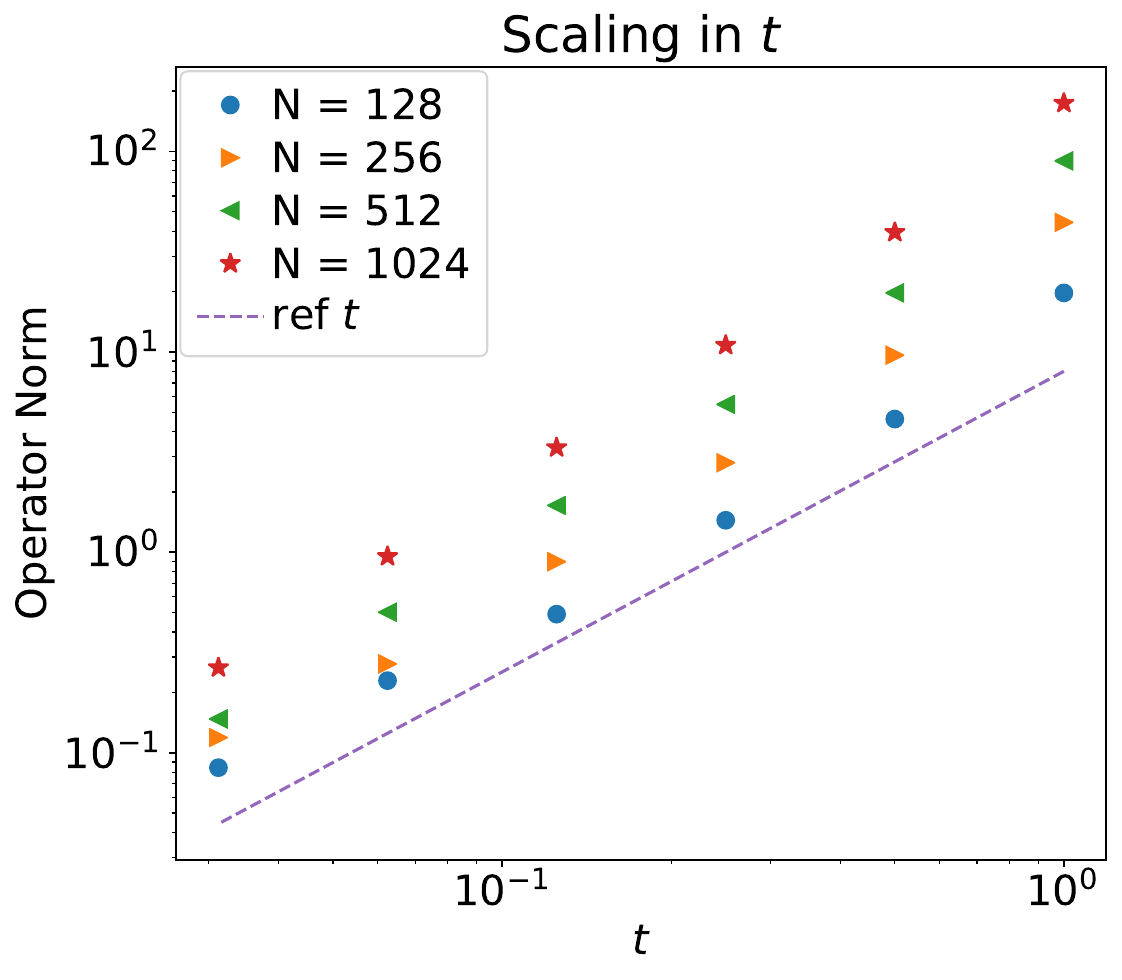}\label{fig:alpha_beta_gamma}}
    \subfloat[Operator norm of the key commutator]{
        \includegraphics[width=.47\textwidth]{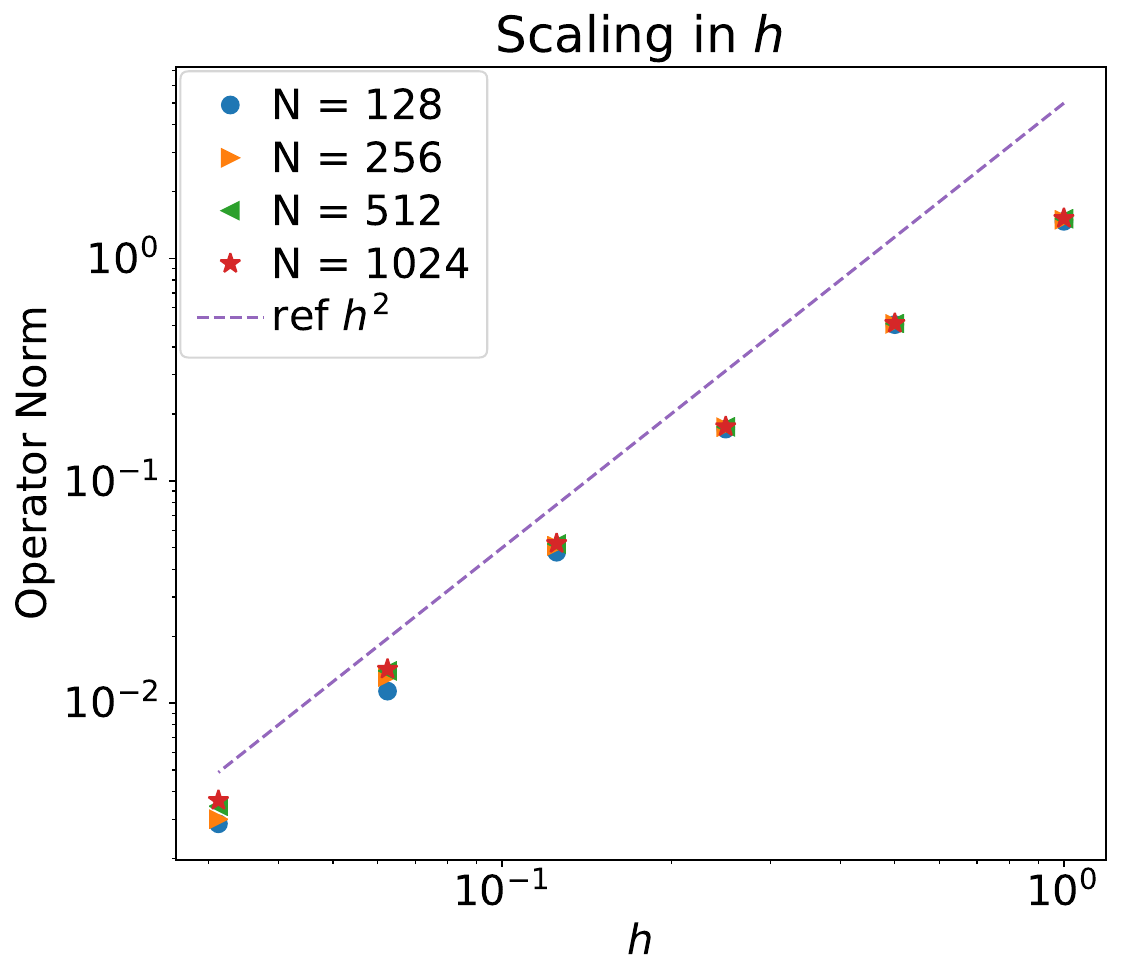} \label{fig:key_comm}}
    \caption{The unbounded Hamiltonian simulation problem considers the Hamiltonian $-\Delta + \cos(x)$, and $A$ and $B$ are the central finite discretization of the operators $-\Delta$ and $\cos(x)$ respectively with periodic boundary conditions using $N$ spatial grids. Left: Plot of the operator norm of the term $[\alpha, [\beta, \gamma(t)]]$ versus the short time $t\in [0, 1]$ for various grid numbers $N$. This serves as evidence for terms becoming uncontrolled when expanding in the Taylor form. Right: Plot of the operator norm of the key commutator as in \cref{lem:bound_commutator_realspace_key}. This demonstrates that the key commutator remains controlled, insensitive to the grid number $N$, in contrast to the term after expanding by Taylor theorems. The numerical evidence agrees with the theoretical result proved in \cref{lem:bound_commutator_realspace_key}. } 
\end{figure}

Despite this observation, it is crucial to emphasize that this does \textit{not} lead to superconvergence. This is primarily because superconvergence not only requires a $t^4$ estimate, but more importantly, it demands the preconstant to be bounded solely by the nature of $B$. When $t$ is not sufficiently small
(in this case, as small as $1/\norm{A}$),
the higher-order terms in the Taylor expansion cease to remain controlled.
To elaborate further, the interaction depicted as $A(t) = -i e^{iAt}B e^{-iAt}$ does not conform to the simple form of \cref{eq:simple_A_expand_in_t}. Instead, it can be expressed as:
\begin{equation}
A(t) = -i B + [A,B] t +i \int_0^t d\sigma e^{i(t-\sigma)A} [A, [A, B]] e^{-i(t-\sigma)A} : = \alpha + \beta t + \gamma(t),
\end{equation}
where $r t^3$ is replaced by $r(t)$, preventing a perfect cancellation of the $t^3$ terms. This leads to terms that are challenging to control, for instance:
\begin{equation} \label{eq:alpha_beta_gamma}
    [\alpha, [\beta, \gamma(t)]]t^3  = [\alpha, [\beta, A(t) - \alpha - \beta t]]t^3
     = [\alpha, [\beta, A(t)]]t^3.
\end{equation}
Here we leverage the fact that\REV{ $[\alpha, [\beta, \alpha]] \sim [V(x), [[\partial_x^2, V(x)], V(x)]] = 0$}, where $\sim$ denotes equality up to an absolute constant.
To achieve superconvergence, \cref{eq:alpha_beta_gamma} needs to be bounded by $C t^4$, necessitating the bounding of $\norm{[\alpha, [\beta, A(t)]]}$ by $Ct$ with some $C$ depending only on $B$ and independent of the number of grids $N$. Unfortunately, this is not the case.
As evidenced by \cref{fig:alpha_beta_gamma}, where the term can be linearly bounded in $t$ but with a constant dependent on the number of grids $N$. Other terms associated with $\gamma(t)$ can be even more challenging to estimate, for example, one needs to prove that $\REV{\sup_{s\in[0,t]}}\norm{[\gamma(s), [\alpha, \gamma(t)]]} \leq C t^3$.

Furthermore, expanding $A(t)$ for one more order does not alleviate these difficulties, and similar challenges arise in the remainder term. The key takeaway is that Taylor expansions or Taylor theorem-style estimates fail to account for the destructive interference resulting from integrals and the cohesive behavior of multiple terms. 
A naive but inaccurate analogy is to consider how $e^{-iH t}$ remains bounded even for unbounded $H$ (the bound is independent of the norm of $H$). However, once one expands it in its Taylor series and attempts to estimate, one ends up with a bound that depends on the norms of $H$, which \REV{is} infinite for unbounded operators \REV{and, after discretization, scales polynomially in the number of spatial grids $N$}. This aligns with the observation of the superconvergence result for the first-order Magnus truncation \cite{AnFangLin2022}. To address such interference and achieve an improved estimate, we pursue the route of pseudodifferential operators in the next section.

\subsection{Rigorous proof of superconvergence}\label{sec:superconv_proof}
In this section, we aim to provide the crucial estimate to establish the superconvergence in the case of the unbounded Hamiltonian simulation in the interaction picture, in particular, $A = \Delta$ represents the Laplacian operator and $B = V(x)$.  
We start by considering an estimate of $I_1$ as defined in \cref{eq:I_1}. \REVV{We reiterate that the goal of this section is to show the simulation error of the unitary operator in the interaction picture. }

The following trivial yet useful fact can come in handy in simplifying the terms.
\begin{equation} \label{eq:simple_fact_comm}
        [e^{iAs} B e^{-iAs}, e^{iAt}C e^{-iAt}]  = e^{iAt}[e^{iA(s-t)}B e^{-iA(s-t)}, C]e^{-iAt}.
\end{equation}
It immediately follows that 
\begin{align*}
    \left[ A(\tau) , [A(s) , A(t)]\right] 
    = & i \left[ e^{iA\tau} B e^{-iA\tau},  e^{iAt}[e^{iA(s-t)}B e^{-iA(s-t)}, B]e^{-iAt} \right]
    \\
    = & i e^{iAt}  \left[ e^{iA(\tau-t)} B e^{-iA(\tau-t)}, [e^{iA(s-t)}B e^{-iA(s-t)}, B] \right]  e^{-iAt}.
\end{align*}
Thus it suffices to show that 
\begin{equation}
   \sup_{\tau, s \in [-h, h]} \norm{ \left[ e^{iA\tau} B e^{-iA\tau}, [e^{iAs}B e^{-iAs}, B] \right]}_{\REVV{\mathcal{L}(L^2)}} \leq C_B h^2,
\end{equation}
where $C_B$ only depends on the potential $V$ and its derivatives. If that is the case, then we have $\|I_1\| \leq C_B h^4$, where the additional two powers of $h$ arise from the double layer integrals over the time interval $[t, t+h]$. This is precisely what we are able to establish in the case of unbounded Hamiltonian simulation, where $A = \Delta$ represents the Laplacian operator and $B = V(x)$ is the potential under certain assumptions. The proof relies on a careful analysis of the pseudodifferential operators, as detailed in Appendix~\ref{app:proof-key-commutator-estimate}.

\begin{lemma}[The key commutator estimate]
\label{lem:bound_commutator_realspace_key}
Let $V \in C^{\infty}(\mathbb{R}^d)$ bounded together with all of its derivatives. Then for every $0 \leq h \leq 1$ we have
\begin{equation} 
\sup_{\tau, s \in [-h, h]} \norm{ \left[ e^{i\Delta\tau} V(x) e^{-i\Delta \tau}, [e^{i \Delta s}V(x) e^{-i\Delta s}, V(x)] \right]}_{\mathcal{L}(L^2)} \leq C_V h^2,
\end{equation}
where $C_V$ is some constant depending only on $V$ and the dimension $d$.
\end{lemma}

\begin{proof}
The proof follows by setting $h_0=1$ in Lemma~\ref{lem:bound_commutator_realspace-general}, proved in Appendix~\ref{app:proof-key-commutator-estimate}. %
\end{proof}

We now address the terms $I_3$ and $I_2$. Although $I_3$ comprises eight terms, they are easy to bound since each term involves at least three layers of time integrals and at least three layers of nested commutators in terms of $A(t)$. Utilizing \cref{eq:simple_fact_comm}, we have
\begin{equation}
    \norm{[A(t), A(s)]}_{\REVV{\mathcal{L}(L^2)}} = \norm{[e^{iA(s-t)}B e^{-iA(s-t)}, B]}_{\REVV{\mathcal{L}(L^2)}}.
\end{equation}
The estimate of the right-hand side is proved in \cite[Lemma 10]{AnFangLin2022} as follows:
\begin{lemma}
\label{lem:bound_commutator_realspace}
For a smooth function $V$ bounded together with all of its derivatives and $0 < h \leq 1$, we have
\begin{equation} 
  \max_{s \in [-h, h]} \norm{[V(x), e^{i s \Delta} V(x) e^{-i s \Delta}]}_{\mathcal{L}(L^2)} \leq C_V h,
\end{equation}
where $C_V$ is some constant depending only on $V$ and the dimension $d$.
\end{lemma}
Moreover, note that $\norm{A(t)}_{\REVV{\mathcal{L}(L^2)}} = \norm{B}_{\REVV{\mathcal{L}(L^2)}}$ depends only on $V$, and three layers of time integrals contribute to $t^3$ or equivalently $h^3$ for $t\in[-h, h]$. Hence, we have $\norm{I_3}_{\REVV{\mathcal{L}(L^2)}} \leq C h^4$, with some constant $C$ depending only on $V$.
Additionally, $I_2 = g(\ad_{\Omega_2})(\ad_{\Omega_2}^3(\dot \Omega_2) )$ follows a similar argument, as $\ad_{\Omega_2}^3(\dot \Omega_2)$ contains at least three layers of time integrals and three layers of nested commutators. We can state the following lemma, which is a direct consequence of the Magnus series expansion. We have provided the proof in \cref{app:pf_lem_g_remainder} for completeness.
\begin{lemma} \label{lem:g_remainder}
For $\Omega_2$ and $g$ defined in \cref{eqn:magnus2_Omega} and \cref{eqn:tilde_A}, we have 
\begin{equation}
    \norm{g(\ad_{\Omega_2})(\ad_{\Omega_2}^3(\dot \Omega_2) )}_{\REVV{\mathcal{L}(L^2)}}\leq C \norm{\ad_{\Omega_2}^3(\dot \Omega_2)}_{\REVV{\mathcal{L}(L^2)}},
\end{equation}
where $C$ is some absolute constant.
\end{lemma}

We comment that this immediately implies that 
\begin{equation}
    \norm{I_2}_{\REVV{\mathcal{L}(L^2)}} \leq 2C \sup_{t \in [-h,h]}\norm{\Omega_2}_{\REVV{\mathcal{L}(L^2)}}\norm{\ad_{\Omega_2}^2(\dot \Omega_2)}_{\REVV{\mathcal{L}(L^2)}}
    \leq C_V h^5,
\end{equation}
where $C_V$ depends only on $V$. The last inequality follows from the fact that $\ad_{\Omega_2}^2(\dot{\Omega}_2)$ contains the terms in $I_1$ and $I_3$ that we have already estimated.

Combining the estimates of $I_1$, $I_2$, and $I_3$, we have the following theorem.
\begin{thm}[Superconvergence of Unbounded Hamiltonian Simulation] \label{thm:superconv} 
For the exact propagator 
 $U_\mathrm{exact}(t,s) = \mathcal{T}  \exp \left( - i\int_s^t H(s) ds\right)$ and the second-order Magnus truncation $U_2(t,s): = e^{\Omega_2(t,s)}$, where $\Omega_2(t,s)$ is defined as \cref{eqn:omega2_t_s_def}, we have
\begin{equation}
    \norm{U_\mathrm{exact}( t_{j+1}, t_j) - U_2(t_{j+1}, t_j)}_{\REVV{\mathcal{L}(L^2)}} \leq C_V h^5,
\end{equation}
for some constant $C_V$ depending only on $V$ and the dimension $d$.
\end{thm}
This implies for the long-time evolution over $[0,T]$, the error is bounded by $C_V T h^4$ so that the algorithm is fourth order.

\REVV{As discussed in \cref{sec:new_rev_wellpose_interaction_picture}, this section mainly focuses on whether our time-dependent Hamiltonian simulation algorithms can accurately simulate the interaction pictured Hamiltonian $H_I(t)$. To recover the Schr\"odinger picture, one needs to further assume the initial condition to be in $H^2$, and extend the bounds to $\mathcal{L}(H^2)$. \cref{lem:bound_commutator_realspace_key} and \cref{lem:bound_commutator_realspace} hold for $\mathcal{L}(H^2)$, because they are both associated with the Weyl quantization of a $S(1)$ symbol, as shown in \cref{lem:commutator-semiclassical-rep}. \cref{lem:g_remainder} holds in the sense of $\mathcal{L}(H^2)$ (see \cref{app:remainder_H2_to_H2}).}

\section{Circuit construction} \label{sec:circuit}
In this section, we first discuss the construction of quantum circuits for dynamical simulation with our algorithm based on the second-order Magnus series truncation. With access to the input model of the time-dependent Hamiltonian in the Schr\"odinger picture or interaction picture, we discuss the complexity of long-time dynamical simulation with second-order Magnus expansion. In particular, we demonstrate that long-time quantum dynamics simulation can be implemented with second-order-Magnus series and superconvergence can be achieved.

\subsection{Input model}\label{sec:input}
The input model refers to the oracle in the quantum circuit giving access to the Hamiltonian of interests. Building an input model for a quantum system is a necessary step before constructing the numerical scheme, for example, quantum signal processing, on quantum computers. Block Encoding is one general framework used to construct an input model on fault-tolerant quantum computer~\cite{GilyenSuLowEtAl2019,sunderhauf2024block}. It has been developed for a number of operators including sparse matrices~\cite{camps2024explicit}, pseudo-differential operator~\cite{li2023efficient} as well as quantum many-body Hamiltonian~\cite{babbush2018encoding,Wan2021exponentially,liu2024efficient,du2024hamiltonian}.

\subsubsection{Input model for $H(t)$}

In this work, we adopt the same input model as used in~\cite{LowWiebe2019}. Assume we are provided with the unitary oracle HAM-T, which encodes the Hamiltonian evaluated at different discrete time steps. Specifically, given a time-dependent Hamiltonian $H(t)$ with $\|H(t)\| \leq \alpha$, two non-negative integers $j, M$, and a time step size $h$, let $\text{HAM-T}_j$ be an $(n_s+n_a+n_m)$-qubit unitary oracle with $n_m = \log_2 M$ such that 
\begin{equation}
    \bra{0^{n_a}}\text{HAM-T}_j\ket{0^{n_a}} =  \sum_{k=0}^{M-1} \ket{k}\bra{k} \otimes \frac{H(jh+kh/M)}{\alpha}. 
\end{equation}
Here $M$ is the number of quadrature points used in numerical quadrature, $j$ is the current local time step, $h$ is the time step size in the time discretization, and $n_s,n_a$ denote the number of the state space qubits and ancilla qubits, respectively.

\subsubsection{Input model for interaction picture}

We assume access to a fast-forwarded Hamiltonian simulation subroutine for the matrix $A$ and the HAM-T oracle for $B(t)$, namely:
\begin{enumerate}
    \item $O_A(s)$ which can fast-forward $e^{i A s}$ for any $s \in \mathbb{R}$. 
    \item $O_B(j)$, which is the HAM-T oracle for $B(t)$ on the interval $[jh,(j+1)h]$, i.e., an $(n_s+n_B+n_m)$-qubit unitary oracle with $n_m = \log_2 M$ and $n_B$ denoting the number of ancilla qubits such that
    \begin{equation}\label{eqn:HAM-T_B_interaction}
        \bra{0^{n_B}} O_B(j) \ket{0^{n_B}} = \sum_{k=0}^{M-1} \ket{k}\bra{k} \otimes \frac{B(jh+kh/M)}{\alpha_B}.
    \end{equation}
    Here $\alpha_B$ is the block-encoding factor such that $\max_{t\in[0,T]}\|B(t)\|\leq \alpha_B$. 
\end{enumerate}

\subsection{Quantum circuit for second-order Magnus expansion}

The unitary $\Tilde{U}_{2}$ defined in  \cref{equ:U_2_approximation} can be implemented on a quantum computer by the linear combination of unitary (LCU)~\cite{ChildsWiebe2012,BerryChildsKothari2015} with HAM-T as an input model and the quantum singular value transformation (QSVT) ~\cite{GilyenSuLowEtAl2019}. One way to further implement \eqref{eqn:tilde_u_2} is to use the quantum circuit to block encodes multiplications of each $\Tilde{U}_2$ operator (See ~\cite{GilyenSuLowEtAl2019} for details). An alternative approach is to block encode the full exponential factor in \cref{eqn:tilde_u_2} and apply QSVT in the end. Both methods require a technique named oblivious amplitude amplification (OAA) to improve the final success probability of post-selection of correct dynamical simulation after block encoding. In this section, we mainly focus on the construction of a short-time evolution operator $\Tilde{U}(t_j,t_{j+1})$ for our algorithm based on the second-order Magnus expansion, and the long-time evolution operator can be constructed afterward.

To implement the double integral in the second-order Magnus expansion, we use a \text{COMP} oracle such that,
\begin{equation} 
\label{equ:comp}
    \text{COMP}\ket{p} \ket{q} \ket{0} = \begin{cases}
    \ket{p} \ket{q} \ket{0} &\text{if $q<p$}\\
    \ket{p} \ket{q} \ket{1}  &\text{\rm{otherwise}},
  \end{cases}
\end{equation}
to control the range of integral. The COMP oracle has been developed and used for a number of applications including state preparation~\cite{sanders2019black}. %
The idea of such COMP oracle lies in comparing the binary representation of two integers from the most significant bits to less important bits~\cite{Oliveira2007QuantumBit}.

\begin{figure}[h]
    \centering
\begin{tikzpicture}
\node[scale=0.53]{
\begin{quantikz}
\lstick[3]{Control} c_1 \ket{0}& \gate{ \text{HAD}} \slice{1} &\qw  &\octrl{1} \slice{2}&\ctrl{4}\slice{3} &\ctrl{1}&\ctrl{4}&\ctrl{2} \slice{4} &\ctrl{2}&\ctrl{4}&\ctrl{1} \slice{5} &\qw  &\ctrl{3} \slice{6}& \gate[1]{S}{} &\gate[1]{\text{HAD}}{}  &\qw \\
c_2 \ket{0^m}& \gate{\otimes_{m} \text{HAD}} & \qw^{p}  &\ctrl{5} &\gate[3]{\text{COMP}}{}&\ctrl{3}&\qw &\qw &\qw&\qw &\ctrl{3}& \qw &\qw &\qw & \gate{\otimes_{m} \text{HAD}}\slice{Final} &\qw  \\
c_3 \ket{0^m}&\gate{\otimes_{m} \text{HAD}} & \qw^{q} &\qw &\qw &\qw &\qw &\ctrl{2}&\ctrl{2}&\qw&\qw &\qw &\qw &\qw &\gate{\otimes_{m} \text{HAD}}& \qw \\ 
\lstick[4]{Ancilla} q_1 \ket{0}&\qw &\qw &\qw &\qw &\qw &\qw &\qw  &\qw&\qw&\qw &\qw & \gate[1]{R_{y}(\theta)}{} &\qw &\qw &\qw \\ 
q_2 \ket{0}&\gate{\text{HAD}} &\qw & \qw &\gate[1]{Z}{-1} &\octrl{2} &\octrl{1} &\octrl{2}&\ctrl{2}&\ctrl{1}&\ctrl{2}&\qw &\qw &\qw &\gate{\text{HAD}} & \qw \\
q_3 \ket{0^a}&\qw &\qw & \qw &\qw &\qw &\swap{1} &\qw &\qw&\swap{1}&\qw &\qw &\qw &\qw &\qw &\qw \\
q_4 \ket{0^a}&\qw &\qw & \gate[2]{\text{HAM-T}_{j}} &\qw  &\gate[2]{\text{HAM-T}_{j}} &\swap{} &\gate[2]{\text{HAM-T}_{j}}&\gate[2]{\text{HAM-T}_{j}}&\swap{} &\gate[2]{\text{HAM-T}_{j}}&\qw &\qw &\qw &\qw &\qw \\
\ket{\psi}&\qw &\qw &\qw &\qw &\qw &\qw  &\qw&\qw &\qw&\qw &\qw &\qw &\qw  & \qw &\qw
\end{quantikz}
};
\end{tikzpicture}
    \caption{Quantum circuit of implementing the block-encoding of the Hamiltonian of the second-order Magnus expansion for general Hamiltonian \cref{equ:second_order_circ}. The short-time evolution operator can then be implemented according to \cref{lem:ham_sim_qsvt} using the circuit here as input block encoding of Hamiltonian. Here HAD is the single qubit Hadamard gate and COMP is a compare oracle defined in \cref{equ:comp}. The $\text{HAM-T}_{j}$ is the input model for the time-dependent Hamiltonian at the time step. For the readability of superscripts and subscripts, we use $a$ and $m$ to represent $n_a$ and $n_m$ respectively in this figure.}
    \label{fig:U_2_circuit}
\end{figure}

\begin{figure}[h]
    \centering
    \begin{tikzpicture}
\node[scale=0.89]{
    \begin{quantikz}
    \lstick[1]{Control}&\ctrl{2}& \qw &\midstick[3,brackets=none]{=}& \qw^{p} &\ctrl{2} &\gate[3]{O_{B}(j)}{} &\ctrl{2}&\qw \\ 
     \lstick[1]{Ancilla}&\gate[2]{\text{HAM-T}_{j}}&\qw && \qw &\qw &\qw &\qw &\qw \\
     \lstick[1]{State}&\qw&\qw& & \gate{O_A(-jh)} &\gate[1]{O_A\left( -\frac{ph}{M}\right)}{}&\qw &\gate[1]{O_A\left( \frac{ph}{M}\right)}{}&\gate[1]{O_A(jh)}{}\\
    \end{quantikz}
};
\end{tikzpicture}
    \caption{Quantum circuit of the HAM-T oracle in the interaction picture Hamiltonian for $H(t)=A+B(t)$.}
    \label{fig:HAM-T}
\end{figure}
 
We implement the Hamiltonian of the short time evolution operator from second-order Magnus expansion in Figure~\ref{fig:U_2_circuit}. The Hamiltonian of the second-order Magnus expansion can be decomposed into the summation of two Hamiltonians, corresponding to first-order and second-order expansion terms of the Magnus series. We use a control qubit $c_1$ and LCU to construct the summation of two terms in \cref{equ:U_2_approximation}. For the first order term, we use LCU approach and $\text{HAM-T}$ oracle to construct a linear combination of $H(t)$ for different $t\in [t_j,t_j+1]$.  The second order term $\frac{1}{2}[H(t_j+\frac{qh}{M}), H(t_j+\frac{ph}{M})]$ is constructed through a LCU and a Pauli Z gate. The LCU technique contributes to the summation of the commutator and $\frac{1}{2}$ and $Z$ gate contribute to the minus sign in the commutator.

To be more precise about how the circuit works, we verify the evolution of wave-state vector $\psi$ under the quantum circuit and prove it is a block encoding of second-order Magnus expansion. 

\begin{lemma}
\label{lem:quadrature_err}
    Quantum circuit illustrated in \cref{fig:U_2_circuit} is a $(2\alpha h,n_s+2 n_m+2 n_a+3,0)$ block encoding of Hamiltonian $H$ derived from second-order Magnus series truncation with time discretization, 
    \begin{equation}
    \label{equ:second_order_circ}
        \sum_{p=0}^{M-1} H(t_j+\frac{ph}{M}) \frac{h}{M}+  \frac{i}{2} \sum_{p=0}^{M-1} \left[ 
  \sum_{q=0}^{p-1} H(t_j+\frac{qh}{M}) \frac{h}{M}, H(t_j+ \frac{
   ph
  }{M}) \right] \frac{h}{M},
    \end{equation}
    defined in \cref{equ:U_2_approximation} for the short time interval $[t_j, t_{j+1}]$.
\end{lemma}

\begin{proof}
	Assume we start with state $\ket{0^{2n_m+2n_a+3}}\ket{\psi}$ where $\psi$ represents a $n_s$-qubit state. The circuit gives 
	\begin{equation}
		\begin{split}
			&\ket{0^{2n_m+2n_a+3}}\ket{\psi} \xrightarrow{Step 1} \frac{1}{2M}\sum_{i=0}^{1}\ket{i} \sum_{p,q=0}^{M-1} \ket{p} \ket{q} \ket{0} \sum_{k=0}^{1}\ket{k}\ket{0^{2n_a}}\ket{\psi}\\
			&\xrightarrow{Step 2} \frac{1}{2\alpha M}\ket{0} \sum_{p,q=0}^{M-1} \ket{p} \ket{q} \ket{0} \sum_{k=0}^{1}\ket{k}\ket{0^{2n_a}} \ket{H(t_j+\frac{ph}{M})\psi}\\
			&\quad + \frac{1}{2M}\ket{1} \sum_{p,q=0}^{M-1} \ket{p} \ket{q} \ket{0} \sum_{k=0}^{1}\ket{k}\ket{0^{2n_a}}\ket{\psi}\\
			&\xrightarrow{Step3} \frac{1}{2 \alpha M}\ket{0} \sum_{p,q=0}^{M-1} \ket{p} \ket{q} \ket{0} \sum_{k=0}^{1}\ket{k}\ket{0^{2n_a}} \ket{H(t_j+\frac{ph}{M})\psi}\\
			&\quad + \frac{1}{2M}\ket{1} \sum_{q<p}^{M-1} \ket{p} \ket{q} \ket{0} \left(\ket{0}\ket{0^{2n_a}}\ket{\psi} - \ket{1}\ket{0^{2n_a}}\ket{\psi}  \right) +\ket{\Psi}_{drop}
   \end{split}
   \end{equation}
   
   Step 1 diffuses the all-zero state to the superposition of states corresponding to the summation of first-order term and second-order term in the Hamiltonian. Step 2 is used to construct the first-order terms in the Hamiltonian. The second-order terms, i.e. the commutators, are constructed from step 3 to step 5 through LCU and a compare oracle. 
   
   \begin{equation}
    \begin{split}
			&\xrightarrow{Step4} \frac{1}{2 \alpha M}\ket{0} \sum_{p,q=0}^{M-1} \ket{p} \ket{q} \ket{0} \sum_{k=0}^{1}\ket{k}\ket{0^{2n_a}} \ket{H(t_j+\frac{ph}{M})\psi}\\
			&\quad + \frac{1}{2 \alpha^2M}\ket{1} \sum_{q<p}^{M-1} \ket{p} \ket{q} \ket{0} \ket{0}\ket{0^{2n_a}}\ket{H(t_j+\frac{qh}{M})H(t_j+\frac{ph}{M})\psi}  \\
    & \quad -\frac{1}{2\alpha^2 M}\ket{1} \sum_{q<p}^{M-1} \ket{p} \ket{q} \ket{0}\ket{1}\ket{0^{2n_a}}\ket{\psi}+\ket{\Psi}_{drop}
    \\
    &\xrightarrow{Step5} \frac{1}{2 \alpha M}\ket{0} \sum_{p,q=0}^{M-1} \ket{p} \ket{q} \ket{0} \sum_{k=0}^{1}\ket{k}\ket{0^{2n_a}} \ket{H(t_j+\frac{ph}{M})\psi} \\
	&\quad + \frac{1}{2\alpha^2 M}\ket{1} \sum_{q<p}^{M-1} \ket{p} \ket{q} \ket{0} \ket{0}\ket{0^{2n_a}}\ket{H(t_j+\frac{qh}{M})H(t_j+\frac{ph}{M})\psi}  \\
   & \quad -\frac{1}{2\alpha^2 M}\ket{1} \sum_{q<p}^{M-1} \ket{p} \ket{q}  \ket{1}\ket{0^{2n_a}}\ket{H(t_j+\frac{ph}{M})H(t_j+\frac{qh}{M})\psi}+\ket{\Psi}_{drop}
		\end{split}
	\end{equation}
	\begin{equation}
		\begin{split}
			&\xrightarrow{Step6} \frac{1}{2\alpha  M}\ket{0} \sum_{p,q=0}^{M-1} \ket{p} \ket{q} \ket{0} \sum_{k=0}^{1}\ket{k}\ket{0^{2n_a}} \ket{H(t_j+\frac{ph}{M})\psi}\\
			&\quad + \frac{h}{2\alpha M}\ket{1} \sum_{q<p}^{M-1} \ket{p} \ket{q} \ket{0} \ket{0}\ket{0^{2n_a}}\ket{H(t_j+\frac{qh}{M})H(t_j+\frac{ph}{M})\psi}   \\  
   & \quad -\frac{h}{2\alpha M}\ket{1} \sum_{q<p}^{M-1} \ket{p} \ket{q} \ket{0}\ket{1}\ket{0^{2n_a}}\ket{H(t_j+\frac{ph}{M})H(t_j+\frac{qh}{M})\psi}+\ket{\Psi}_{drop}
   \\
			&\xrightarrow{End} \frac{1}{2\alpha M}\ket{0} \ket{0^{n_m}} \ket{0^{n_m}} \ket{0} \ket{0}\ket{0^{2a}} \sum_{p=0}^{M-1}\ket{H(t_j+\frac{ph}{M})\psi}\\
			&\quad + \frac{ih}{4\alpha M^2}\ket{0} \ket{0^{n_m}} \ket{0^{n_m}} \ket{0} \sum_{q<p}^{M-1}\ket{0}\ket{0^{2n_a}}\ket{H(t_j+\frac{qh}{M})H(t_j+\frac{ph}{M})\psi}  \\ 
    &\quad -\frac{ih}{4\alpha M^2}\ket{0} \ket{0^{n_m}} \ket{0^{n_m}} \ket{0} \sum_{q<p}^{M-1} \ket{0}\ket{0^{2n_a}}\ket{H(t_j+\frac{ph}{M})H(t_j+\frac{qh}{M})\psi}+\ket{\Psi}_{drop}
		\end{split}
	\end{equation}
	After each step, we use the notation $\ket{\Psi}_{drop}$ to denote the dropped term after post-selection. Although written in ket notation, we note that the state is not normalized and the notation may represent a different state in each step. For the post-selection, we measure all the control qubits and ancilla qubits. Only when all the measurement results are zero, we accept the output of the quantum circuit. In step 6, we use a control-rotation gate to adjust the coefficients and the rotation angle is,
	\begin{equation}
		\theta = \text{arccos}(\alpha h).
	\end{equation}
	The detailed derivation verifies the circuit is a $(2\alpha h,2n_m+2n_a+3,0)$ block encoding of time discretization of second-order Magnus expansion.
\end{proof}
\begin{rem}
The construction of COMP oracle requires two additional ancilla qubits but can be uncomputed, and we do not include the two qubits here.
\end{rem}

\subsection{Long-time complexity of second-order Magnus expansion}

In order to estimate the complexity of long-time quantum dynamics simulation through the second-order Magnus expansion, we start the section by recalling the following lemma~\cite{GilyenSuLowEtAl2019} (see also \cite[Lemma 2]{AnFangLin2022}).
\begin{lemma}[Time-independent Hamiltonian simulation via QSVT and OAA]\label{lem:ham_sim_qsvt}
    Let $\epsilon \in (0,1)$, $t = \Omega(\epsilon)$ and let $U$ be an $(\alpha,n_a,0)$-block-encoding of a time-independent Hamiltonian $H$. 
    Then a unitary $V$ can be implemented such that $V$ is a $(1,n_a+2,\epsilon)$-block-encoding of $e^{-i t H}$, with $\Or\left(\alpha t + \log(1/\epsilon)\right)$ uses of $U$, its inverse or controlled version,  $\Or\left(n_a(\alpha t + \log(1/\epsilon))\right)$ elementary gates and $\mathcal{O}(1)$ additional ancilla qubits. 
\end{lemma}

\cref{lem:quadrature_err} and \cref{lem:ham_sim_qsvt} contribute to the estimation of dynamical simulation cost with second-order Magnus expansion, summarized in the \cref{thm:long_time_cost}. 

\begin{thm}[Long-time complexity of Second-order Magnus expansion]
\label{thm:long_time_cost}
    Let the Hamiltonian $H(s)$ satisfies $\|H(s)\| \leq \alpha$ for all $0\leq s \leq T$. If
    \begin{equation}
        \norm{U_\mathrm{exact}( t_{j+1}, t_j) - U_2(t_{j+1}, t_j)} \leq C_H h^{1+\theta}.
    \end{equation}
    for a non-negative real number $\theta$ and a constant $C_H$ which might depend on $H$. 
    Then for any $0 < \epsilon < 1, T > \epsilon$, our algorithm can implement an operation $\exp(\tilde{\Omega}_2$ such that $\|W - U(T,0)\| \leq \epsilon$ with failure probability $\mathcal{O}(\epsilon)$ and the following cost: 
    \begin{enumerate}
        \item $\mathcal{O}\left(\alpha T + \frac{C_{H}^{1/\theta}T^{1+1/\theta}}{\epsilon^{1/\theta}}\log\left(\frac{C_{H}T}{\epsilon}\right)\right)$ uses of $\text{HAM-T}_j$, its inverse or controlled version, 
        \item $\mathcal{O}\left(\alpha T + \frac{C_{H}^{1/\theta}T^{1+1/\theta}}{\epsilon^{1/\theta}}\log\left(\frac{C_{H}T}{\epsilon}\right)\right)$ uses of \text{COMP} oracle or controlled version.
        \item $\mathcal{O}\left(\left(n_a + \log\left(\frac{\max_{s\in [0,T]} \norm{H'(s)}T^{1-1/\theta}}{C_{H}^{1/\theta}\epsilon^{1-1/\theta}}\right)\right)\left(\alpha T + \frac{C_{H}^{1/\theta}T^{1+1/\theta}}{\epsilon^{1/\theta}}\log\left(\frac{C_{H}T}{\epsilon}\right)\right)\right)$ one- or two-qubit gates, 
        \item $\mathcal{O}\left(\log\left(\frac{\max_{s\in [0,T]} \norm{H'(s)}T}{C_{H}\epsilon}\right)\right)$  ancilla qubits. 
    \end{enumerate}
    
\end{thm}
\begin{proof}
We start with~\cref{fig:U_2_circuit} which is a $(2\alpha h, n_s+2n_a+2n_m+3,0)$-block-encoding of second order Magnus series in~\cref{equ:second_order_circ}, 
\begin{equation}
    \sum_{p=0}^{M-1} H(t_{j}+\frac{ph}{M}) \frac{h}{M}+  \frac{i}{2} \sum_{p=0}^{M-1} \left[ 
  \sum_{q=0}^{p-1} H(t_j+\frac{qh}{M}) \frac{h}{M}, H(t_j+ \frac{
  ph
  }{M}) \right] \frac{h}{M}
\end{equation}
with five query to $\text{HAM-T}_{j}$ oracle, one query to \text{COMP} oracle, $O(n_m)$ single-qubit gates and two controlled-SWAP gates. According to Lemma~\ref{lem:ham_sim_qsvt}, a $(1, 2n_a+2n_m+5,\delta)$-block-encoding of $\tilde{U}_{2}(t_{j+1},t_j)$ can be implemented by QSVT, with $\mathcal{O} (2\alpha h+\text{log}(1/\delta))$ uses of the block encoding for second-order Magnus expansion, $\mathcal{O} ((n_s+2n_a+2n_m+3)(2\alpha h+\text{log}(1/\delta)))$ elementary gates and $\mathcal{O}(1)$ additional ancilla qubits. By setting 
\begin{equation}
 M= \Or\left(\frac{\max_{s\in [0,T]} \norm{H'(s)}}{C_{H}h^{\theta-1}} \right),
\end{equation}
the implementation is a $(1, n_s+2n_a+2n_m+5,\delta')$-block-encoding of $U_{2}(t_{j+1},t_{j})$, where
\begin{equation}
    \delta' = \delta+2C_{H}h^{\theta+1}.
\end{equation}
Let $V_j$ denote the circuit constructed in \cref{fig:U_2_circuit} and let $n_b=2n_a+2n_m+5$, then we have for $W_j=\bra{0}_{b} V_j\ket{0}_{b}$,
\begin{equation}
    \norm{W_j-U_2(t_{j+1},t_{j})}\leq \delta'.
\end{equation}
With $\norm{W_j}\leq 1$ and $\norm{U(t_{j+1},t_{j})}$, we have 
\begin{equation}
    \norm{\Pi_{j=0}^{L-1}W_j-U(T,0)}\leq L\delta'.
\end{equation}
The success probability of evolving the state $\psi$, then can be estimated as 
\begin{equation}
    \norm{(\bra{0}_{b}\otimes I_{n_s}) \Pi_{j=0}^{L-1}V_{j} \ket{0}_{b}\ket{\psi}}
\end{equation}
where $I_{n_s}$ denotes the identity operator on the $n_s$ qubits. Since $\norm{W_j-U_2(t_{j+1},t_{j})}\leq \delta'$, the success probability of each step, is thus bounded by
\begin{equation}
    1-(1-\delta')^2< 2\delta'.
\end{equation}
After $L$ steps, the failure probability is bounded by
\begin{equation}
    1-(1-2\delta')^{L} \leq 2L \delta'.
\end{equation}
To bound the total error below $\Or(\epsilon)$, we require
\begin{equation}
    L\delta'=   L\delta +2\frac{C_{H}T^{\theta+1}}{L^{\theta}} \leq 3 \epsilon.
\end{equation}
By bounding the first and second terms on the left side by $\epsilon$ and $2\epsilon$ respectively, it suffices to choose
\begin{equation}
    L=\frac{C_{H}^{1/\theta}T^{1+1/\theta}}{\epsilon^{1/\theta}}, \ \delta= \frac{\epsilon^{1+1/\theta}}{C_{H}^{1/\theta}T^{1+1/\theta}},
\end{equation}
and correspondingly
\begin{equation}
 M= \Or\left( \frac{\max_{s\in [0,T]} \norm{H'(s)}T^{1-1/\theta}}{C_{H}^{1/\theta}\epsilon^{1-1/\theta}}\right).
\end{equation}
\end{proof}
\begin{rem}
    We note that the multi-qubit controlled $\text{HAM-T}_{j}$ is counted as a single query to $\text{HAM-T}_{j}$. For a clear illustration of the circuit logic, we put the qubit $c_1$ to the top and there could be additional cost due to long-range control for controlled-SWAP gate or controlled-Z gate. However, in real implementation, the cost can be removed as the qubit $c_1$ can be moved to the middle of $c_3$ and $q_1$. 
\end{rem}

We summarize the query complexity of the input model for simulating long-time quantum dynamics with second-order Magnus expansion for both general $H$ and the interaction picture with superconvergence in \cref{tab:cost_compare1}. Note that although the scaling of the derivative bound and superconvergence appear similar, it is important to note that for unbounded Hamiltonian simulation, $C_V$ in the superconvergence result is $N$ independent, while $C'_H$ depends on $N$ polynomially, where $N$ is the number of spatial grids.
    \begin{table}[h!]
    \centering
    \begin{tabular}{c|c}
      General $H$ & $\mathcal{O}\left(\alpha T + \frac{C_\mathrm{comm}^{1/2}T^{3/2}}{\epsilon^{1/2}}\log\left(\frac{C_\mathrm{comm}T}{\epsilon}\right)\right)$\\
      \hline
      General $H$ with derivative bound & $\mathcal{O}\left(\alpha T + \frac{(C_H')^{1/4}T^{5/4}}{\epsilon^{1/4}}\log\left(\frac{C_{H}'T}{\epsilon}\right)\right)$\\ 
      \hline
      Superconvergence & $\mathcal{O}\left(\alpha_{B} T + \frac{C_{V}^{1/4}T^{5/4}}{\epsilon^{1/4}}\log\left(\frac{C_{V}T}{\epsilon}\right)\right)$\\
    \end{tabular}

    \caption{Comparison of query complexities of long time dynamics simulation with second order Magnus series based on different assumptions of Hamiltonian $H$. For general Hamiltonian $H$, query complexities are measured by the number of queries to the input of the time-dependent Hamiltonian. For the interaction picture, the query complexity is measured with respect to the queries to the oracle $e^{-iAt}$ or input model of $B$.}
    \label{tab:cost_compare1}
\end{table}

\section{Conclusion and discussion}\label{sec:conclusion}

In this work, we develop a quantum algorithm based on the second-order Magnus expansion for general time-dependent Hamiltonian simulation problems. It has commutator scaling in the high-precision regime, and the cost is logarithmically dependent on the derivative of the underlying Hamiltonian, as our algorithm leverages a quadrature rule with many quadrature points and then implements it using an LCU-type circuit. Our algorithm functions similarly to a non-interpolatory Magnus integrator, meaning it avoids the errors typically induced by approximating the integral with only a few quadrature points. We further show that the algorithm is in general of second order. If one allows a polynomial dependence on the derivative, the algorithm can be of fourth order. Surprisingly, when the Hamiltonian simulation is taken to be the interaction picture of the unbounded Hamiltonian simulation for regular potentials, the algorithm exhibits a fourth-order superconvergence, where the error preconstant becomes only dependent on the potential $V$, making the overall gate cost only logarithmic dependent on $N$. This distinguishes our algorithm from other (classical) Magnus integrators that introduce polynomial derivative dependence of the Hamiltonian.

So far, the direct implementation of the Magnus expansion for general time-dependent Hamiltonians has been developed only for expansions up to the second order. In theory, the approach of truncating the Magnus series to a finite order and then implementing many quadrature points for the integrals via a suitable LCU-type circuit could be applied to arbitrarily high-order truncations. However, challenges can arise in both the circuit construction and the superconvergence proof. Exploring both challenges will be interesting future directions.

For the circuit, we implement the second-order Magnus expansion with a compare oracle to post-select the terms satisfying $q<p$ in \cref{equ:U_2_approximation}. Each register representing $p$ or $q$ requires $\mathcal{O}\left(\log(\frac{T}{\epsilon}) \right)$ qubits and the commutator is constructed with LCU techniques requiring one additional ancilla qubit. Direct generalization is less efficient as $k$-th order Magnus series requires $k$ qubits for the composition of $k$ layers commutator and $\mathcal{O}(k)$ qubits for registers controlling the ordering of time register. Construction of a compression gadget to deal with the two issues for optimal implementation will be left for future directions.  

\REV{We expect that truncating the Magnus series at arbitrarily high orders can eventually achieve the same cost scaling in $t$ and $\epsilon$ as truncating the high-order Dyson series, specifically $\widetilde{\mathcal{O}}(t)$ and $\mathcal{O}(\log(\epsilon))$. However, the truncation error of Magnus series involves a nested commutator type preconstant, which is smaller than the constant in Dyson scaling, as it depends on the operator norm of $H(t)$ itself. Nevertheless, the circuit construction for high-order Magnus methods is expected to be more complicated, similar to that of the high-order Dyson series, which would require a control unit for time clocking. Therefore, while higher-order methods can offer significantly improved asymptotic scaling, lower-order algorithms might be more practical to implement.} 
Although we can confidently conjecture that the superconvergence behavior extends to generalizations of our algorithms for high-order Magnus expansions, proving this for higher-order expansions presents substantial challenges. One clear reason is that we cannot rely on series expansions, such as the Taylor expansion. As we have seen, the superconvergence proof for the second-order Magnus expansion already involves 11 terms. Nevertheless, our proof highlights the cancellation mechanism through semiclassical analysis and categorizes the terms into a few groups, providing insights for the generalization to the arbitrary higher-order scenario. 

\REV{In principle, the semiclassical analysis argument used here to prove superconvergence can be applied to any $A, B$ that are elements in the Lie algebra
$$\mathcal{L}: = \{\sum_{k=0}^n y_k(x) \partial_x^k, n \in \mathbb{Z}_+, y_k \in S(1), \text{ for } k = 1, \cdots, n\}.$$ Thus, the error estimates for such general cases should not pose significant issues. However, the catch is that for an efficient quantum algorithm for such an unbounded Hamiltonian, we convert to the interaction picture, which would require $A$ to be fast-forwardable, while not all elements in this algebra can be fast-forwarded.
A broader exploration of quantum algorithms in the context of superconvergence remains an interesting topic for future study.}
Another interesting direction would be to seek other systems that can make better use of this type of commutator scaling and explore whether the superconvergence behavior exists for other physically motivated Hamiltonians. \REV{Though semiclassical analysis is primarily designed for unbounded operators, it may still be possible to apply these tools to bounded Hamiltonians due to the connection between matrices and quantization on a quantized torus for continuous phase-space functions~\cite{BornsWeilFang2022}.}

\REVV{
Finally, we remark that the superconvergence result is rigorously proven only for $S(1)$ potentials. However, the Coulomb potential and other unbounded potentials do not belong to $S(1)$. Investigating whether similar superconvergence behavior may occur for special initial conditions in the state-dependent case remains an interesting direction for future research. Additionally, our superconvergence result holds for continuous spatial degrees of freedom. While such analysis is typically considered for spectral methods, it would also be interesting to extend this proof to other discretizations, such as finite difference methods for finite $N$.
}

\section*{Acknowledgments}

The authors would like to thank Andr\'{a}s Vasy for discussions on semiclassical microlocal analysis that led to the proof of Lemma~\ref{lem:bound_commutator_realspace-general}\REV{, and Israel Michael Sigal and Avy Soffer for their helpful comments on the connection to operator calculus. The authors would also like to thank the anonymous referee for the insightful suggestion to carefully check the domain for unbounded Hamiltonian simulation in the interaction picture. } We also thank the Institute for Pure and Applied Mathematics (IPAM) for hosting all authors as long-term participants during the semester-long program “Mathematical and Computational Challenges in Quantum Computing” in Fall 2023, during which this work was initiated.

\section*{Declarations}

\textbf{Funding}. This material is based upon work supported by National Science Foundation via the grant DMS-2347791, the U.S. Department of Energy, Office of Science, Accelerated Research in Quantum Computing Centers, Quantum Utility through Advanced Computational Quantum Algorithms, grant no. DE-SC0025572, and Ralph E. Powe Junior Faculty Enhancement Award (D.F.), DMREF Award No. 1922165, Simons Targeted Grant Award No.
896630 and Willard Miller Jr. Fellowship (D.L.). This work was also supported by the National Science Foundation via Award No. 1953987, and partially supported by the Simons Investigator Award in Mathematics, which is a grant from the Simons Foundation (825053), as well as by the U.S. Department of Energy, Office of Science, National Quantum Information Science Research Centers, Quantum Systems Accelerator (R.S.).

\noindent
\textbf{Financial interests}. The authors declare that they have no competing financial interests.

\noindent
\textbf{Conflict of interest statement and Data availability statement}.
On behalf of all authors, the corresponding author states that there is no conflict of interest and data sharing is not applicable to this article as no datasets were generated or analyzed during the current study.

\appendix

\section{Proof of the key commutator estimate 
}
\label{app:proof-key-commutator-estimate}

The goal of this section is to prove Lemma~\ref{lem:bound_commutator_realspace-general}, from which Lemma~\ref{lem:bound_commutator_realspace} follows as a special case. The proof uses standard results from semiclassical analysis \cite[Chapter~4]{zworski2022semiclassical}, and the result is an immediate consequence of these general results. We also give some necessary background for the ease of the reader. All functions will be complex-valued.

For any $m \in \mathbb{R}$, we begin by defining the symbol class $S_{3n}(\langle p \rangle^m)$, where $\langle p \rangle = (1 + |p|^2)^{\frac{1}{2}}$, and the reader should consult \cite[Chapter~4]{zworski2022semiclassical} and \cite[Chapter~2]{Martinez2002} for more information. Here the subscript $3n$ corresponds to three copies of $\mathbb{R}^n$, which we will denote as $\mathbb{R}^n_{x}$, $\mathbb{R}^n_{y}$, $\mathbb{R}^n_{p}$, for reasons that will become clear below. For some $h_0 > 0$ (which we will think of as fixed), the definition of $S_{3n}(\langle p \rangle^m)$ is now as follows:
\begin{equation}
\label{eq:S3n-def}
\begin{split}
    S_{3n}(\langle p \rangle^m) &:= \{a \in C^{\infty}(\mathbb{R}^n_{x} \times \mathbb{R}^n_{y} \times \mathbb{R}^n_{p} \times (0,h_0]_h) : \\
    &\sup_{\mathbb{R}^{3n} \times (0,h_0]_h} | \langle p \rangle^{-m} (\partial_x^{\alpha} \partial_y^{\beta} \partial_p^{\gamma} a )(x,y,p, h)| \leq C_{\alpha,\beta,\gamma} < \infty \},
\end{split}
\end{equation}
that is the space of all smooth functions on $\mathbb{R}^{3n} \times (0,h_0]_h$, which are bounded together with all derivatives with respect to $x,y,p$ after scaling by $\langle p \rangle^{-m}$, and these bounds are uniform in $h \in (0,h_0]$. It is traditional in semiclassical analysis to hide the dependence of symbols $a \in S_{3n}(\langle p \rangle^m)$ on $h$ for the ease of notation, and we will adopt this convention and simply write $a(x,y,p)$ with dependence on $h$ implicit. Similarly, we also define the symbol class 
\begin{equation}
\label{eq:S2n-def}
\begin{split}
    S_{2n}(\langle p \rangle^m) &:= \{a \in C^{\infty}(\mathbb{R}^n_{x}  \times \mathbb{R}^n_{p} \times (0,h_0]_h) : \\
    &\sup_{\mathbb{R}^{2n} \times (0,h_0]_h} |\langle p \rangle^{-m} (\partial_x^{\alpha} \partial_p^{\gamma} a )(x,p, h)| \leq C_{\alpha,\gamma} < \infty \}.
\end{split}
\end{equation}
In Eq.~\eqref{eq:S3n-def} and Eq.~\eqref{eq:S2n-def}, $\alpha, \beta, \gamma$ are multi-indices. For the special case $m=0$, we will denote the corresponding symbol classes as simply $S_{3n}(1)$ and $S_{2n}(1)$. Let us now introduce semiclassical pseudodifferential operators defined by the (semiclassical) quantization of symbols in $S_{3n}(\langle p \rangle^{m})$. For any $h \in (0,h_0]$ and $a \in S_{3n}(\langle p \rangle^{m})$, we define the quantization of $a$ to be
\begin{equation}
\label{eq:quantization-S3n}
    \text{Op}_{h}(a) := \frac{1}{(2 \pi h)^n} \int_{\mathbb{R}^n} a(x,y,p) e^{\frac{i}{h}(x-y) \cdot p} \; dp.
\end{equation}
We also define the Weyl quantization of symbols in $S_{2n}(\langle p \rangle^{m})$: if $a \in S_{2n}(\langle p \rangle^{m})$, then it gives rise to a semiclassical pseudodifferential operator
\begin{equation}
\label{eq:quantization-S2n-Weyl}
    \text{Op}_{h}^{\text{w}}(a) := \frac{1}{(2 \pi h)^n} \int_{\mathbb{R}^n} a \left(\frac{x+y}{2},p \right) e^{\frac{i}{h}(x-y) \cdot p} \; dp.
\end{equation}
Formally the expressions in Eq.~\eqref{eq:quantization-S3n} and Eq.~\eqref{eq:quantization-S2n-Weyl} define the Schwartz kernels of the corresponding pseudodifferential operators, which are \text{oscillatory integral distributions}.

As preparation for our proof, we will also need to restate some of the calculations done in the proof of \cite[Lemma~10]{AnFangLin2022}, but using semiclassical quantization. In particular, we note the following result, the proof of which follows using similar calculations as in \cite{AnFangLin2022} (we omit the proof):

\begin{lemma}[Weyl quantization formulas]
\label{lem:weyl-quant-formulas}
Let $V \in C^{\infty}(\mathbb{R}^n_{x})$, such that $V$ is bounded with all derivatives. Let us also assume that $a \in S_{2n}(1)$, and $h \in (0, h_0]$. Then we have the following:
\begin{enumerate}[(i)]
    \item We have the equality $[\Delta, \text{Op}_{h}^{\text{w}}(a)] = \frac{2i}{h} \text{Op}_{h}^{\text{w}} \left( \sum_{i=1}^{n} p_i \frac{\partial a}{\partial x_i} \right)$, where $\sum_{i=1}^{n} p_i \frac{\partial a}{\partial x_i} \in S_{2n} (\langle p \rangle)$.
    \item For all $s \in \mathbb{R}$, we have $e^{ish \Delta} V e^{-ish \Delta} = \text{Op}_{h}^{\text{w}}(V(x - 2ps))$, where $(V(x-2ps)) \in S_{2n}(1)$, and does not depend on $h$.
\end{enumerate}
\end{lemma}

Using the above facts, the proof of Lemma~\ref{lem:bound_commutator_realspace-general} will follow using two standard results in semiclassical analysis. We begin by stating these results below.

\begin{lemma}[Semiclassical commutator, {\cite[Theorem~2.7.4, Remark~2.7.6]{Martinez2002}}]
\label{lem:semiclassical-commutator}
Let $a, b \in S_{2n}(1)$, and let $\text{Op}_{h}^{\text{w}}(a)$ and $\text{Op}_{h}^{\text{w}}(b)$ denote their Weyl quantizations. Then
\begin{enumerate}[(i)]
    \item There exists $c, \tilde{c} \in S_{2n}(1)$ such that $\text{Op}_{h}^{\text{w}}(a) \text{Op}_{h}^{\text{w}}(b) = \text{Op}_{h}^{\text{w}}(c)$, and we can write $c$ as
    \begin{equation}
    \begin{split}
        &c = ab + \frac{h}{2i} \{a, b\} + h^2 \tilde{c}, \\
        &\{a, b\} := \sum_{j=1}^{n} \left( \frac{\partial a}{\partial p_j} \frac{\partial b}{\partial x_j} - \frac{\partial a}{\partial x_j} \frac{\partial b}{\partial p_j} \right).
    \end{split}
    \end{equation}
    \item For the commutator $[\text{Op}_{h}^{\text{w}}(a), \text{Op}_{h}^{\text{w}}(b)]$, there exists a symbol $c \in S_{2n}(1)$ such that 
    \begin{equation}
        [\text{Op}_{h}^{\text{w}}(a), \text{Op}_{h}^{\text{w}}(b)] = \frac{h}{i} \text{Op}_{h}^{\text{w}} \left(  \{a,b\} \right) + h^3 \text{Op}_{h}^{\text{w}}(c).
    \end{equation}
\end{enumerate}
\end{lemma}

\begin{lemma}[Calder\'{o}n-Vaillancourt, {\cite[Theorem~2.8.1]{Martinez2002}}]
\label{lem:calderon-vaillancourt}
If $a \in S_{3n}(1)$, then $\text{Op}_{h}(a)$ defines a continuous linear map from $L^2(\mathbb{R}^n_{y})$ to $L^2(\mathbb{R}^n_{x})$. In particular, we have
\begin{equation}
\label{eq:Calderon-Vaillancourt}
    \norm{\text{Op}_{h}(a)}_{L^2 \rightarrow L^2} \leq C_n \left( \sum_{|\alpha| \leq M_n} \sup_{(x,y,p) \in \mathbb{R}^{3n}} |\partial^{\alpha} a (x,y,p)| \right),
\end{equation}
where the constants $C_n$ and $M_n$ depend only on $n$.
\end{lemma}
We next prove another important lemma that we will need:

\begin{lemma}
\label{lem:commutator-semiclassical-rep}
Let $V \in C^{\infty}(\mathbb{R}^n_{x})$ such that $V$ is bounded together with all its derivatives. Then we have the following:
\begin{enumerate}[(i)]
    \item If $s \in [-h_0, h_0]$, then there exists $a \in S_{2n}(1)$, such that $[e^{i \Delta s}V e^{-i\Delta s}, V] = s \text{Op}_{|s|}^{\text{w}}(a)$.
    \item Assume that $s,\tau \in [-h,h]$ for some $h \in (0,h_0]$. Let $s = rh, \; \tau = qh$. Then there exists $a_{r,q} \in S_{2n}(1)$, depending continuously on the parameters $r,q \in [-1,1]$, such that
    \begin{equation}
        [e^{i \Delta \tau}V e^{-i\Delta \tau}, [e^{i \Delta s}V e^{-i\Delta s}, V]] = h^2 \text{Op}_{h}^{\text{w}}(a_{r,q}). 
    \end{equation}
\end{enumerate}
\end{lemma}

\begin{proof}
For this proof, note that the assumptions on $V$ ensure that $V \in S_{2n}(1)$.
\begin{enumerate}[(i)]
    \item For $s=0$, we have $[e^{i \Delta s}V e^{-i\Delta s}, V] = 0$; so let us assume that $0 \neq s = th_0$, for some $t$. Then from Lemma~\ref{lem:weyl-quant-formulas}(ii), we have
    \begin{equation}
        e^{is \Delta} V e^{-is \Delta} = e^{ith_0 \Delta} V e^{-ith_0 \Delta} = \text{Op}_{h_0}^{\text{w}}(V(x - 2pt)).
    \end{equation}
    Now, for $t > 0$, a simple variable change $pt \mapsto p$ shows that we have 
    \begin{equation}
        e^{is \Delta} V e^{-is \Delta} = \text{Op}_{h_0}^{\text{w}}(V(x - 2pt)) = \text{Op}_{t h_0}^{\text{w}}(V(x - 2p)) = \text{Op}_{s}^{\text{w}}(V(x - 2p)),
    \end{equation}
    where $V(x-2p) \in S_{2n}(1)$. The importance of the above equation is that we have managed to make the symbol $V(x-2p)$ independent of $t$. Next using the fact that $V = \text{Op}_{s}^{\text{w}}(V)$, it immediately follows from Lemma~\ref{lem:semiclassical-commutator} that there exists $c \in S_{2n}(1)$ such that
    \begin{equation}
    \begin{split}
        [e^{i \Delta s}V e^{-i\Delta s}, V] &= [\text{Op}_{s}^{\text{w}}(V(x - 2p)), \text{Op}_{s}^{\text{w}}(V)] = \frac{s}{i} \text{Op}_{s}^{\text{w}}(b) + s^3 \text{Op}_{s}^{\text{w}}(c) \\
        &=  s \text{Op}_{s}^{\text{w}} \left( \frac{b}{i} + s^2 c \right), \\
        b(x,p) &:= - 2 (\nabla V)(x-2p) \cdot (\nabla V)(x).
    \end{split}
    \end{equation}
    It follows that $b \in S_{2n}(1)$, as $V \in S_{2n}(1)$, and thus $(b/i + s^2 c) \in S_{2n}(1)$ also. Finally, the case $t<0$ is similar and follows after performing the variable change $-pt \mapsto p$.

    \item We have $e^{is \Delta} V e^{-is \Delta} = \text{Op}_{h}^{\text{w}}(V(x - 2pr))$, and $e^{i \tau \Delta} V e^{-i \tau \Delta} = \text{Op}_{h}^{\text{w}}(V(x - 2pq))$, from Lemma~\ref{lem:weyl-quant-formulas}(ii). Next we express the commutator $[e^{is \Delta} V e^{-is \Delta}, V]$ using Lemma~\ref{lem:semiclassical-commutator}(ii) as
    \begin{equation}
        [e^{is \Delta} V e^{-is \Delta}, V] = [\text{Op}_{h}^{\text{w}}(V(x - 2pr)), \text{Op}_{h}^{\text{w}}(V)] = h \text{Op}_{h}^{\text{w}} \left( \frac{b_r}{i} + h^2 c_r \right),
    \end{equation}
    where $b_r, c_r \in S_{2n}(1)$, with $b_r(x,p) := -2r (\nabla V)(x-2pr) \cdot (\nabla V)(x)$, and also note that $b_r, c_r$ depend continuously on $r$. Define $\tilde{b}_r := b_r/i + h^2 c_r$.

    We can now calculate the commutator $[e^{i \tau \Delta} V e^{-i \tau \Delta}, [e^{is \Delta} V e^{-is \Delta}, V]]$ the same way:
    \begin{equation}
    \begin{split}
        &[e^{i \tau \Delta} V e^{-i \tau \Delta}, [e^{is \Delta} V e^{-is \Delta}, V]] = [\text{Op}_{h}^{\text{w}}(V(x-2pq)), h \text{Op}_{h}^{\text{w}}(\tilde{b}_r)] \\
        &= h [\text{Op}_{h}^{\text{w}}(V(x-2pq)), \text{Op}_{h}^{\text{w}}(\tilde{b}_r)] = h^2 \left( \text{Op}_{h}^{\text{w}}(d_{r,q}/i + h^2 f_{r,q}) \right),
    \end{split}
    \end{equation}
    where $d_{r,q}, f_{r,q} \in S_{2n}(1)$ depend continuously on $r,q$, and
    \begin{equation}
        d_{r,q}(x,p) := -2q (\nabla V)(x-2pq) \cdot (\nabla_x \tilde{b}_r)(x,p) + (\nabla V)(x-2pq) \cdot (\nabla_p \tilde{b}_r)(x,p).
    \end{equation}
    This completes the proof.
\end{enumerate} 
\end{proof}

Note that part (i) of Lemma~\ref{lem:commutator-semiclassical-rep} can also be found in the proof of \cite[Lemma~10]{AnFangLin2022}, but we have derived the same fact using semiclassical tools. We are now ready to prove the main result of this appendix:

\begin{lemma}
\label{lem:bound_commutator_realspace-general}
Let $h_0 > 0$, and $V \in C^{\infty}(\mathbb{R}^n)$ bounded together with all of its derivatives. Then for every $0 \leq h \leq h_0$ we have
\begin{equation} 
\sup_{\tau, s \in [-h, h]} \norm{ \left[ e^{i\Delta\tau} V(x) e^{-i\Delta \tau}, [e^{i \Delta s}V(x) e^{-i\Delta s}, V(x)] \right]}_{L^2 \rightarrow L^2} \leq C_V h^2,
\end{equation}
where $C_V$ is some constant depending only on $V$, $n$, and $h_0$.
\end{lemma}
\begin{proof}
Fix $s,\tau \in [-h,h]$, and suppose $s=hr,\; \tau=hq$. Using Lemma~\ref{lem:commutator-semiclassical-rep}(ii), we know that there exists $a_{r,q} \in S_{2n}(1)$, depending continuously on $r,q \in [-1,1]$, such that $[e^{i \Delta \tau}V e^{-i\Delta \tau}, [e^{i \Delta s}V e^{-i\Delta s}, V]] = h^2 \text{Op}_{h}^{\text{w}}(a_{r,q})$. Now by Lemma~\ref{lem:calderon-vaillancourt}, we have
\begin{equation}
\begin{split}
    &\norm{[e^{i \Delta \tau}V e^{-i\Delta \tau}, [e^{i \Delta s}V e^{-i\Delta s}, V]]}_{L^2 \rightarrow L^2} = \norm{h^2 \text{Op}_{h}^{\text{w}}(a_{r,q})}_{L^2 \rightarrow L^2} \\
    &\leq C_n h^2 \left( \sum_{|\alpha| \leq M_n} \sup_{(x,p) \in \mathbb{R}^{2n}} |\partial^{\alpha} a_{r,q} (x,p)|  \right),
\end{split}
\end{equation}
where the constants $C_n, M_n$ depend on the dimension $n$ only. Now note that by continuity of $a_{r,q}$ as a function of $r,q$, we know that the functions $\sup_{(x,p) \in \mathbb{R}^{2n}} |\partial^{\alpha} a_{r,q} (x,p)|$ are also continuous functions of $r,q \in [-1,1]$, for every multi-index $\alpha$. Finally, we take supremum over $s,\tau \in [-h,h]$ of both sides of the above equation to get
\begin{equation}
\begin{split}
    &\sup_{\tau, s \in [-h, h]} \norm{[e^{i \Delta \tau}V e^{-i\Delta \tau}, [e^{i \Delta s}V e^{-i\Delta s}, V]]}_{L^2 \rightarrow L^2} \\
    & \leq C_n h^2 \left( \sum_{|\alpha| \leq M_n} \sup_{r, q \in [-1, 1]} \left(\sup_{(x,p) \in \mathbb{R}^{2n}} |\partial^{\alpha} a_{r,q} (x,p)| \right) \right) = C(n,V,h_0) h^2.
\end{split}
\end{equation}
The constant $C(n, V, h_0)$ only depends on $n$, the potential function $V$, and implicitly on the parameter $h_0$ (which we had fixed at the beginning).
\end{proof}

\section{Proof of \cref{lem:g_remainder}} \label{app:pf_lem_g_remainder}
\begin{proof}[Proof of \cref{lem:g_remainder}]
The proof of this lemma is fairly straightforward following \cite[Lemma 5.1]{HochbruckLubich2003}, where it is shown that $g(z)$ has a Fourier transform $\hat g \in L^1(\mathbb{R})$ defined as
\begin{equation}
    g(ix) = \int_\mathbb{R} \hat g(\xi) e^{i\xi x} \, d\xi,
\end{equation}
and its norm is bounded by some absolute constant
\begin{equation}\label{eq:hat_g_L1}
\norm{\hat g}_{L^1(\mathbb{R})} \leq 2\pi \norm{g}_{L^2(i\mathbb{R})}^{1/2}\norm{g'}_{L^2(i\mathbb{R})}^{1/2}.
\end{equation}
Plugging in the operators, one has
\begin{equation}\label{eqn:omega_2_fourier}
    g(\ad_{\Omega_2})(\ad_{\Omega_2}^3(\dot \Omega_2)) =
    \int_\mathbb{R} \hat g(\xi) e^{\ad(\xi \Omega_2)} \ad_{\Omega_2}^3(\dot \Omega_2)  \, d\xi
    =  \int_\mathbb{R} \hat g(\xi) e^{\xi \Omega_2} \ad_{\Omega_2}^3(\dot \Omega_2) e^{-\xi \Omega_2} \, d\xi. 
\end{equation}
Note that $\Omega_2$ is anti-Hermitian so that its exponential is unitary. Taking the norm of \cref{eqn:omega_2_fourier}, we have 
\begin{align}
         \norm{g(\ad_{\Omega_2})(\ad_{\Omega_2}^3(\dot \Omega_2) )}_{\REVV{\mathcal{L}(L^2)}} \leq \norm{\hat g}_{L^1(\mathbb{R})} \norm{\ad_{\Omega_2}^3(\dot \Omega_2)}_{\REVV{\mathcal{L}(L^2)}}.  
\end{align}

\end{proof}

\section{\REVV{Proof that the remainder maps $H^2$ to $H^2$
}} \label{app:remainder_H2_to_H2}

\REVV{
In this section, we prove that 
\[
g(\ad_{\Omega_2})(\ad_{\Omega_2}^3(\dot \Omega_2))
\] 
maps from $H^2$ to $H^2$.
}

\REVV{
Note that $\Omega_2$ corresponds to the Weyl quantization of an $S(1)$ symbol (up to a constant $i$), as follows from its definition in \cref{eqn:magnus2_Omega} and the fact that $A(t) = -i H_I(t)$ also corresponds to the Weyl quantization of an $S(1)$ symbol (up to a constant $i$). So is $\dot{\Omega}(t)$ by its definition in \cref{eqn:magnus2_dOmega}, as $S(1)$ is stable under Moyal product.
Note that the term of interest can be expressed as
\begin{equation} \label{eq:g_remainder_integral_xi}
g(\ad_{\Omega_2})(\ad_{\Omega_2}^3(\dot \Omega_2)) = \int_\mathbb{R} \hat g(y) e^{y \Omega_2} \ad_{\Omega_2}^3(\dot \Omega_2) e^{-y \Omega_2}  dy,
\end{equation}
This propels us to consider $e^{y \Omega_2} \ad_{\Omega_2}^3(\dot \Omega_2) e^{-y \Omega_2}$, which also corresponds to the Weyl quantization of a $S(1)$ symbol. This follows from the Egorov theorem, which in particular implies that  
for symbols $a, b \in S(1)$, 
\begin{equation}\label{eq:egorov}
    e^{\frac{i }{h}\op_h(b)}\op_h(a)e^{-\frac{i }{h}\op_h(b)}=\op_h (a_1),
\end{equation} 
for some symbol $a_1 \in S(1)$. In fact, one can derive more information about the expression of $a_1$, but it is not essential here. For the proof of this theorem, see, e.g., \cite[Theorem 11.1]{zworski2022semiclassical}. So we have that $e^{y \Omega_2} \ad_{\Omega_2}^3(\dot \Omega_2) e^{-y \Omega_2}$ maps from $H^s$ to $H^s$ for any $s\in \mathbb{Z}_+$.
This is again because we can apply \cite[Theorem 8.10]{zworski2022semiclassical} with $m_1 = 1$ and $m_2 = \langle \xi \rangle^s$, as discussed in \cref{eq:H2_to_H2-forS1}.
Notice that the norm in fact can be upper bounded independent of $y$. Consider $\psi \in H^2$,
\begin{equation}\label{eq:H2_norm_w_y}
\begin{aligned}
    \norm{e^{y \Omega_2} \ad_{\Omega_2}^3(\dot \Omega_2) e^{-y \Omega_2} \psi }_{H^2} 
    & \leq \norm{e^{y \Omega_2}}_{\mathcal{L}(H^2)} \norm{\ad_{\Omega_2}^3(\dot \Omega_2)}_{\mathcal{L}(H^2)}\norm{ e^{-y \Omega_2}}_{\mathcal{L}(H^2)} \norm{ \psi }_{H^2}
\end{aligned}
\end{equation}
which is bounded independent of $y$, because $i \Omega_2$ is a self-adjoint operator from $H^2$ to $H^2$ as discussed in \cref{sec:new_rev_wellpose_interaction_picture}.
}

\REVV{Recall the definition of the Sobolev norm
\begin{equation}
  \norm{\phi(x)}_{H^2_x} = \norm{(1+|\xi|^2) \left( \mathcal{F} \phi \right)(\xi) } _{L^2_\xi},
\end{equation}
 where $\mathcal{F}$ denotes the Fourier transform in the variable $x$. We need to show for any $\psi(x) \in H^2_x$,
 \begin{equation}\label{eq:sobolev_S_to_prove}
   S : =   \norm{(1+|\xi|^2) \int_\mathbb{R} dy \int_{\mathbb{R}^d} dx e^{-i\xi \cdot x}  \hat g(y) e^{y \Omega_2} \ad_{\Omega_2}^3(\dot \Omega_2) e^{-y \Omega_2} \psi(x)   } _{L^2_\xi}
 \end{equation}
 is bounded.
Note that 
\begin{equation}
 \norm{(1+|\xi|^2)\int_{\mathbb{R}^d} dx e^{-i\xi \cdot x} e^{y \Omega_2} \ad_{\Omega_2}^3(\dot \Omega_2) e^{-y \Omega_2} \psi(x)}_{L^2_\xi}
\end{equation}
is bounded. This is because the Fourier transform is an isometry from $H^2(\mathbb{R}^d)$ to $L^2( \mathbb{R}^d; (1+|\xi|^2) d\xi)$. In other words, using the shorthand notation we have
\begin{equation}
   w_y(x) : = e^{y \Omega_2} \ad_{\Omega_2}^3(\dot \Omega_2) e^{-y \Omega_2} \psi(x) \in H^2_x,
\end{equation}
and its $H^2$ norm is bounded independent of y, as shown in \cref{eq:H2_norm_w_y}. Hence $( (1+|\xi|^2) \mathcal{F}w_y \in L^2_\xi$ with a norm independent of $y$.
Applying the Minkowski's inequality for integrals to \cref{eq:sobolev_S_to_prove}, we have
\begin{equation}
\begin{aligned}
     S  & \leq  \int_\mathbb{R} dy \norm{(1+|\xi|^2)  \int_{\mathbb{R}^d} dx e^{-i\xi \cdot x}  \hat g(y) e^{y \Omega_2} \ad_{\Omega_2}^3(\dot \Omega_2) e^{-y \Omega_2} \psi(x)   } _{L^2_\xi}
     \\
     & \leq \int_\mathbb{R} dy |\hat g(y)| \norm{(1+|\xi|^2) \mathcal{F}w_y} _{L^2_\xi} = \norm{\hat g}_{L^1} \norm{(1+|\xi|^2) \mathcal{F}w_y} _{L^2_\xi},
\end{aligned} 
\end{equation}
which is bounded. This completes the proof.
}

\REVV{
In fact, we remark that it is possible to directly write an exact error representation of the second-order Magnus expansion in terms of only nested commutator of the two and three layers without the terms involving \eqref{eq:g_remainder_integral_xi}, which would make it much easier to demonstrate the $H^2$ to $H^2$ boundedness of the error terms.
}

\REV{
\section{Numerical Comparison}
Note that we have proven all results rigorously. Nevertheless, to further validate our theoretical findings, we present additional numerical results for the Schrödinger equation in the interaction picture.
For simplicity, we consider the following Hamiltonian
\begin{equation}
H= -\Delta + V(x), \quad V(x) = \cos(x), \quad x\in [-\pi,\pi]
\label{eqn:ham_lap_v}
\end{equation}
with periodic boundary conditions. Here $A$ corresponds to the discretized $-\Delta$ using a second order finite difference scheme, and $B$ the discretized $V(x)$, respectively.}

\REV{
We first present the convergence rate with respect to the time step $h$ for various spatial grid numbers $N$. Both our quantum Magnus integrator (q-Mag) and the classical Magnus integrator (c-Mag)~\cite{BlanesCasasOteoRos2009} are considered. In the superconvergence case, our second-order algorithm becomes fourth order, and hence we compare it with the fourth-order classical Magnus integrator. \cref{fig:quantum-classical-Mag} plots the operator norm error versus the time step $h$ in the log-log scale. Here the values of $h$ are chosen as $2^{-1}, 2^{-2}, \cdots, 2^{-5}$ and $M$ is fixed to be a sufficiently large number $2^{17}h$. The system is simulated until the final time $t = 0.5$ and the number of spatial discretization $N$ is taken as $32, 64, 128, 256$.
It can be seen that both our algorithm and the 4th-order classical Magnus integrator exhibits the fourth order convergence in $h$. However, the error of classical Magnus integrator error grows with respect to the increase of $N$ while our quantum Magnus integrator remains insensitive.
}

\REV{
Next we compare our algorithm based on the second-order Magnus expansion with qHOP~\cite{AnFangLin2022}. The parameters are chosen the same as before. In the superconvergence case, as shown in \cref{fig:qMag_qHOP}, the algorithm of this work (q-Mag) exhibits fourth-order convergence with an error preconstant independent of the spatial grid number $N$, while our previous work (qHOP) exhibits second-order convergence. To achieve the same error, q-Mag requires much fewer time steps compared to qHOP.
\begin{figure}
    \centering
    \includegraphics[width=.6\textwidth]{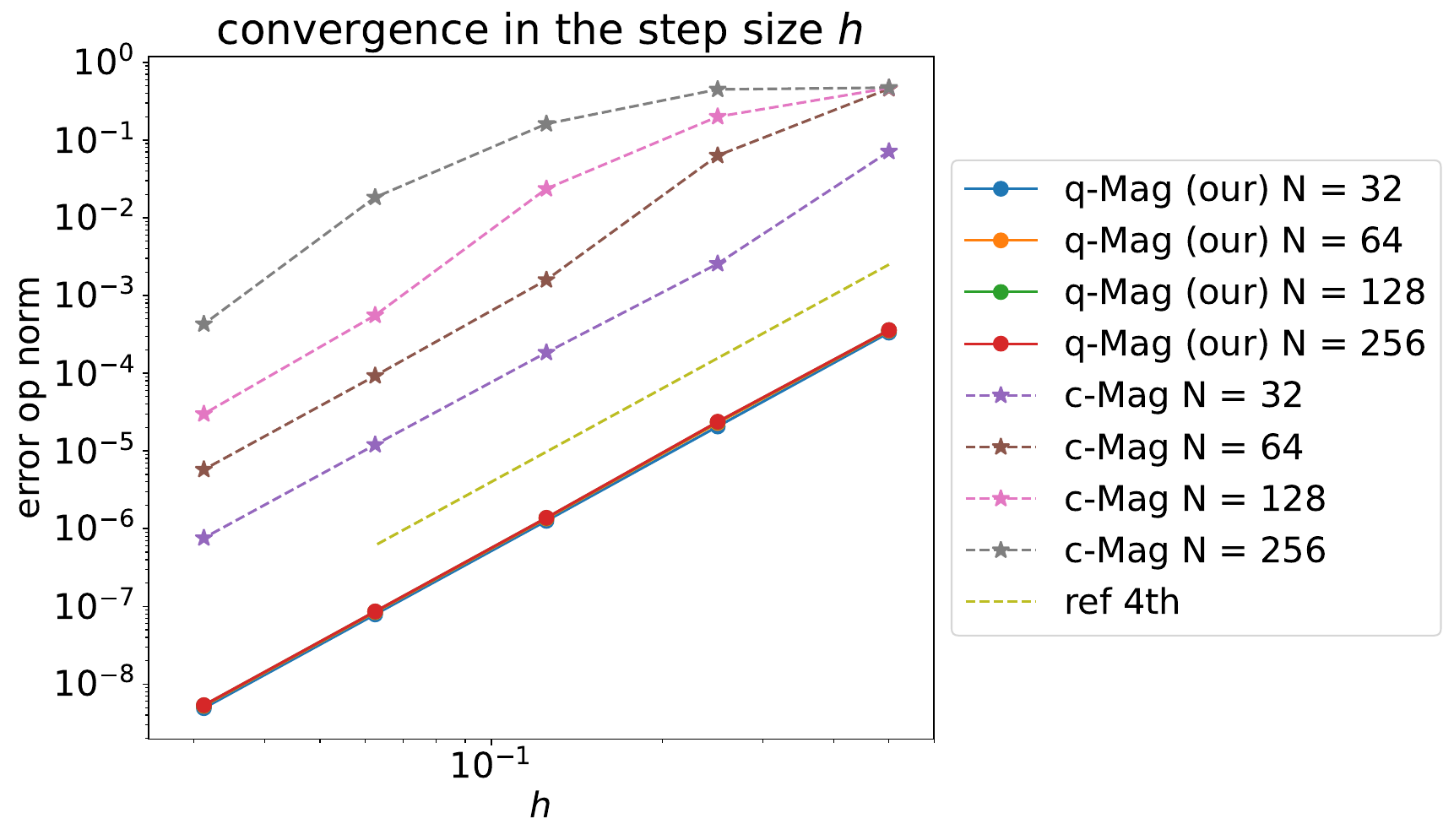}
    \caption{Log-log plot of the errors in the operator norm for various time step sizes $h$. The spatial discretization is finite difference. Both our algorithm (q-Mag) and the fourth-order classical Magnus integrator (c-Mag) exhibit fourth-order convergence. However, the error of our algorithm is smaller than that of the fourth-order classical Magnus integrator, and does not grow as the number of the grid points in spatial discretization $N$ increases. The reference line demonstrates the asymptotic scaling.}
    \label{fig:quantum-classical-Mag}
\end{figure} 
\begin{figure}
    \centering
    \includegraphics[width=.6\textwidth]{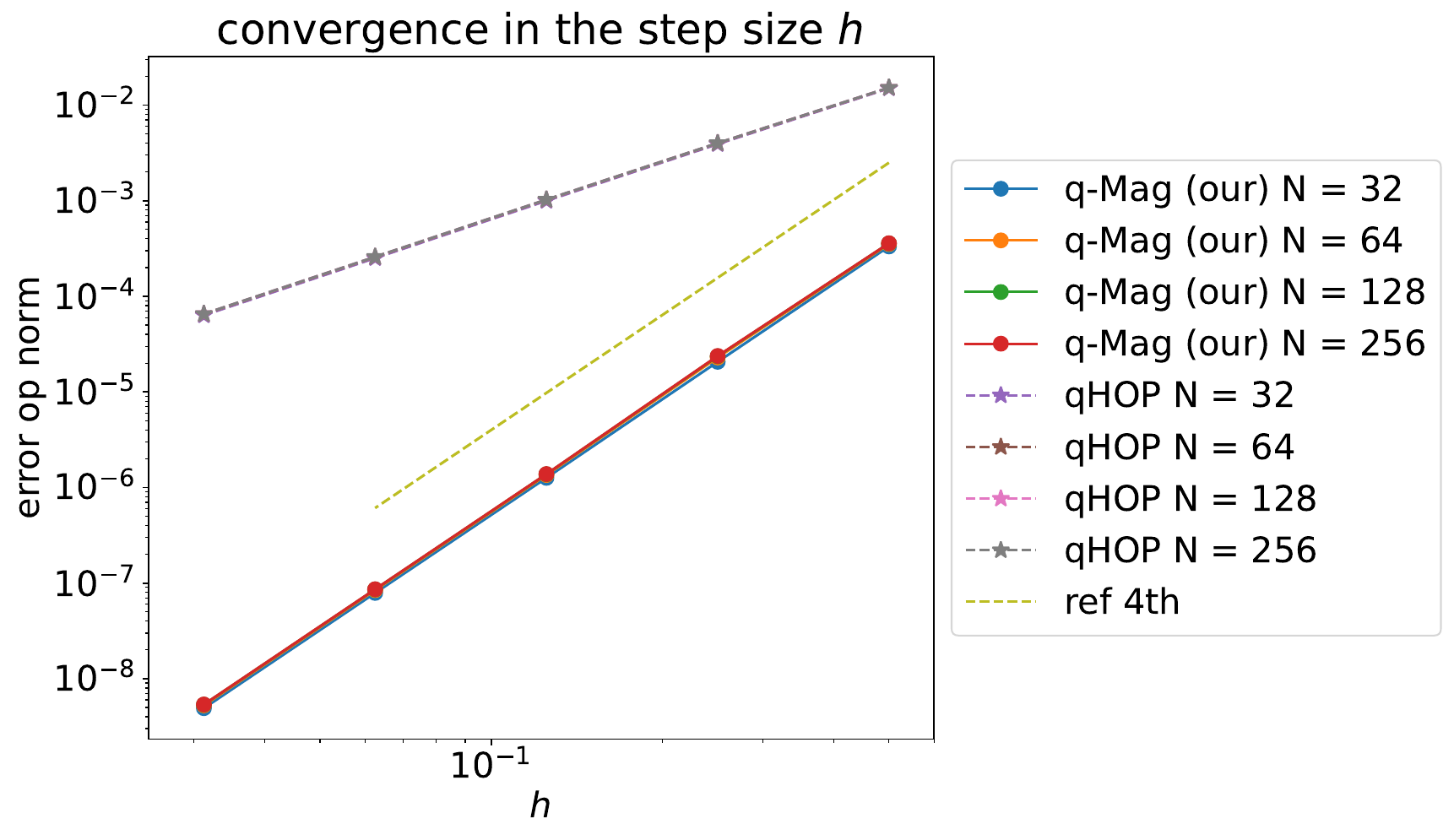}
    \caption{Log-log plot of the errors in the operator norm for various time step sizes $h$ of both this algorithm (q-Mag) and qHOP. The convergence of q-Mag is much faster than qHOP, which implies that to achieve the same precision the number of time steps needed in q-Mag is much fewer than that of qHOP.}
    \label{fig:qMag_qHOP}
\end{figure} 
}

\begin{spacing}{0}
\bibliographystyle{unsrt}
\bibliography{magnus}

\begin{thebibliography}{10}

\bibitem{Feynman1982}
Richard~P Feynman.
\newblock {Simulating physics with computers}.
\newblock {\em International Journal of Theoretical Physics}, 21(6):467--488,
  1982.

\bibitem{FarhiGoldstoneGutmannEtAl2000}
Edward Farhi, Jeffrey Goldstone, Sam Gutmann, and Michael Sipser.
\newblock Quantum computation by adiabatic evolution.
\newblock 2000.

\bibitem{AlbashLidar2018}
T.~Albash and D.~A. Lidar.
\newblock Adiabatic quantum computation.
\newblock {\em Rev. Mod. Phys.}, 90:015002, 2018.

\bibitem{LowChuang2017}
Guang~Hao Low and Isaac~L. Chuang.
\newblock Optimal {H}amiltonian simulation by quantum signal processing.
\newblock {\em Phys. Rev. Lett.}, 118:010501, 2017.

\bibitem{LowChuang2019}
Guang~Hao Low and Isaac~L. Chuang.
\newblock Hamiltonian simulation by qubitization.
\newblock {\em Quantum}, 3:163, Jul 2019.

\bibitem{ChildsSuTranEtAl2020}
Andrew~M. Childs, Yuan Su, Minh~C. Tran, Nathan Wiebe, and Shuchen Zhu.
\newblock Theory of trotter error with commutator scaling.
\newblock {\em Phys. Rev. X}, 11:011020, 2021.

\bibitem{Bassman2022PRL}
Lindsay Bassman~Oftelie, Katherine Klymko, Diyi Liu, Norm~M. Tubman, and
  Wibe~A. de~Jong.
\newblock Computing free energies with fluctuation relations on quantum
  computers.
\newblock {\em Phys. Rev. Lett.}, 129:130603, Sep 2022.

\bibitem{KivlichanWiebeBabbushEtAl2017}
I.~D. Kivlichan, N.~Wiebe, R.~Babbush, and A.~Aspuru-Guzik.
\newblock Bounding the costs of quantum simulation of many-body physics in real
  space.
\newblock {\em J. Phys. A Math. Theor.}, 50:305301, 2017.

\bibitem{Somma2015}
Rolando~D. Somma.
\newblock Quantum simulations of one dimensional quantum systems, 2015.

\bibitem{KivlichanMcCleanWiebeEtAl2018}
I.D. Kivlichan, J.~McClean, N.~Wiebe, C.~Gidney, A.~Aspuru-Guzik, G.K.-L. Chan,
  and R.~Babbush.
\newblock {Quantum Simulation of Electronic Structure with Linear Depth and
  Connectivity}.
\newblock {\em Phys. Rev. Lett.}, 120(11):110501, 2018.

\bibitem{AnFangLin2021}
Dong An, Di~Fang, and Lin Lin.
\newblock Time-dependent unbounded {H}amiltonian simulation with vector norm
  scaling.
\newblock {\em {Quantum}}, 5:459, may 2021.

\bibitem{SuBerryWiebeEtAl2021}
Yuan Su, Dominic~W Berry, Nathan Wiebe, Nicholas Rubin, and Ryan Babbush.
\newblock Fault-tolerant quantum simulations of chemistry in first
  quantization.
\newblock {\em PRX Quantum}, 2(4):040332, 2021.

\bibitem{ChildsLengEtAl2022}
Andrew~M. Childs, Jiaqi Leng, Tongyang Li, Jin-Peng Liu, and Chenyi Zhang.
\newblock Quantum simulation of real-space dynamics, 2022.

\bibitem{RubinBerryKononovMaloneEtAl2023}
Nicholas~C. Rubin, Dominic~W. Berry, Alina Kononov, Fionn~D. Malone, Tanuj
  Khattar, Alec White, Joonho Lee, Hartmut Neven, Ryan Babbush, and Andrew~D.
  Baczewski.
\newblock Quantum computation of stopping power for inertial fusion target
  design, 2023.

\bibitem{TongAlbertMccleanPreskillSu2022}
Yu~Tong, Victor~V. Albert, Jarrod~R. McClean, John Preskill, and Yuan Su.
\newblock Provably accurate simulation of gauge theories and bosonic systems.
\newblock {\em {Quantum}}, 6:816, September 2022.

\bibitem{AbrahamsenTongBaoSuWiebe2023}
Nilin Abrahamsen, Yu~Tong, Ning Bao, Yuan Su, and Nathan Wiebe.
\newblock Entanglement area law for one-dimensional gauge theories and bosonic
  systems.
\newblock {\em Phys. Rev. A}, 108:042422, Oct 2023.

\bibitem{MizrahiNeyenhuisJohnsonEtAl2013}
J.~Mizrahi, B.~Neyenhuis, K.~G. Johnson, W.~C. Campbell, C.~Senko, D.~Hayes,
  and C.~Monroe.
\newblock Quantum control of qubits and atomic motion using ultrafast laser
  pulses.
\newblock {\em Applied Physics B}, 114(1-2):45–61, Nov 2013.

\bibitem{NielsenDowlingGuEtAl2006}
Michael~A Nielsen, Mark~R Dowling, Mile Gu, and Andrew~C Doherty.
\newblock Optimal control, geometry, and quantum computing.
\newblock {\em Phys. Rev. A}, 73(6):062323, 2006.

\bibitem{DongPetersen2010}
Daoyi Dong and Ian~R Petersen.
\newblock Quantum control theory and applications: a survey.
\newblock {\em IET Control Theory \& Applications}, 4(12):2651--2671, 2010.

\bibitem{LowWiebe2019}
G.~H. Low and N.~Wiebe.
\newblock Hamiltonian simulation in the interaction picture.
\newblock 2019.

\bibitem{RajputRoggeroWiebe2021}
Abhishek Rajput, Alessandro Roggero, and Nathan Wiebe.
\newblock Hybridized methods for quantum simulation in the interaction picture,
  2021.

\bibitem{EcksteinMansurogluCzarnikZhuEtAl2023}
Timo Eckstein, Refik Mansuroglu, Piotr Czarnik, Jian-Xin Zhu, Michael~J.
  Hartmann, Lukasz Cincio, Andrew~T. Sornborger, and Zoë Holmes.
\newblock Large-scale simulations of floquet physics on near-term quantum
  computers, 2023.

\bibitem{KuwaharaMoriSaito2016}
Tomotaka Kuwahara, Takashi Mori, and Keiji Saito.
\newblock Floquet–magnus theory and generic transient dynamics in
  periodically driven many-body quantum systems.
\newblock {\em Annals of Physics}, 367:96–124, April 2016.

\bibitem{ZhangLengLi2021}
Chenyi Zhang, Jiaqi Leng, and Tongyang Li.
\newblock Quantum algorithms for escaping from saddle points.
\newblock {\em {Quantum}}, 5:529, August 2021.

\bibitem{LiuSuLi2023}
Yizhou Liu, Weijie~J. Su, and Tongyang Li.
\newblock On quantum speedups for nonconvex optimization via quantum tunneling
  walks.
\newblock {\em Quantum}, 7:1030, June 2023.

\bibitem{LengHickmanLiWu2023}
Jiaqi Leng, Ethan Hickman, Joseph Li, and Xiaodi Wu.
\newblock Quantum hamiltonian descent, 2023.

\bibitem{ChildsSu2019}
Andrew~M Childs and Yuan Su.
\newblock Nearly optimal lattice simulation by product formulas.
\newblock {\em Phys. Rev. Lett.}, 123(5):050503, 2019.

\bibitem{SahinogluSomma2020}
Burak {\c{S}}ahino{\u{g}}lu and Rolando~D Somma.
\newblock {Hamiltonian simulation in the low energy subspace}.
\newblock 2020.

\bibitem{SuHuangCampbell2021}
Yuan Su, Hsin-Yuan Huang, and Earl~T. Campbell.
\newblock Nearly tight {T}rotterization of interacting electrons.
\newblock {\em {Quantum}}, 5:495, July 2021.

\bibitem{ZhaoZhouShawEtAk2021}
Qi~Zhao, You Zhou, Alexander~F. Shaw, Tongyang Li, and Andrew~M. Childs.
\newblock Hamiltonian simulation with random inputs, 2021.

\bibitem{FangTres2023}
Di~Fang and Albert Tres~Vilanova.
\newblock Observable error bounds of the time-splitting scheme for
  quantum-classical molecular dynamics.
\newblock {\em SIAM J. Numer. Anal.}, 61(1):26--44, 2023.

\bibitem{BornsWeilFang2022}
Yonah Borns-Weil and Di~Fang.
\newblock Uniform observable error bounds of trotter formulae for the
  semiclassical schrödinger equation, 2022.

\bibitem{ZengSunJiangZhao2022}
Pei Zeng, Jinzhao Sun, Liang Jiang, and Qi~Zhao.
\newblock Simple and high-precision hamiltonian simulation by compensating
  trotter error with linear combination of unitary operations, 2022.

\bibitem{GongZhouLi2023}
Weiyuan Gong, Shuo Zhou, and Tongyang Li.
\newblock A theory of digital quantum simulations in the low-energy subspace.
\newblock {\em arXiv preprint arXiv:2312.08867}, 2023.

\bibitem{LowSuTongTran2023}
Guang~Hao Low, Yuan Su, Yu~Tong, and Minh~C. Tran.
\newblock Complexity of implementing trotter steps.
\newblock {\em PRX Quantum}, 4:020323, May 2023.

\bibitem{ZhaoZhouChilds2024}
Qi~Zhao, You Zhou, and Andrew~M. Childs.
\newblock Entanglement accelerates quantum simulation, 2024.

\bibitem{ChenXuZhaoYuan2024}
Boyang Chen, Jue Xu, Qi~Zhao, and Xiao Yuan.
\newblock Error interference in quantum simulation, 2024.

\bibitem{WatkinsWiebeRoggeroLee2022}
Jacob Watkins, Nathan Wiebe, Alessandro Roggero, and Dean Lee.
\newblock Time dependent hamiltonian simulation using discrete clock
  constructions, 2022.

\bibitem{ZhukRobertsonBravyi2023}
Sergiy Zhuk, Niall Robertson, and Sergey Bravyi.
\newblock Trotter error bounds and dynamic multi-product formulas for
  hamiltonian simulation, 2023.

\bibitem{AftabAnTrivisa2024}
Junaid Aftab, Dong An, and Konstantina Trivisa.
\newblock Multi-product hamiltonian simulation with explicit commutator
  scaling, 2024.

\bibitem{HuyghebaertDeRaedt1990}
J.~Huyghebaert and H.~De~Raedt.
\newblock Product formula methods for time-dependent {S}chr\"{o}dinger
  problems.
\newblock {\em J. Phys. A}, 23(24):5777--5793, 1990.

\bibitem{WiebeBerryHoyerEtAl2010}
N.~Wiebe, D.~Berry, P.~H{\o}yer, and B.~C. Sanders.
\newblock Higher order decompositions of ordered operator exponentials.
\newblock {\em J. Phys. A}, 43(6):065203, 2010.

\bibitem{BerryChildsCleveEtAl2015}
D.~W. Berry, A.~M. Childs, R.~Cleve, R.~Kothari, and R.~D. Somma.
\newblock Simulating {Hamiltonian} dynamics with a truncated {Taylor} series.
\newblock {\em Phys. Rev. Lett.}, 114:090502, 2015.

\bibitem{KieferovaSchererBerry2019}
Mária Kieferová, Artur Scherer, and Dominic~W. Berry.
\newblock Simulating the dynamics of time-dependent hamiltonians with a
  truncated dyson series.
\newblock {\em Physical Review A}, 99(4), Apr 2019.

\bibitem{BerryChildsSuEtAl2020}
D.~W. Berry, A.~M. Childs, Y.~Su, X.~Wang, and N.~Wiebe.
\newblock Time-dependent {H}amiltonian simulation with $l^{1}$-norm scaling.
\newblock {\em Quantum}, 4:254, 2020.

\bibitem{PoulinQarrySommaEtAl2011}
David Poulin, Angie Qarry, Rolando Somma, and Frank Verstraete.
\newblock Quantum simulation of time-dependent {H}amiltonians and the
  convenient illusion of {H}ilbert space.
\newblock {\em Phys. Rev. Lett.}, 106(17):170501, 2011.

\bibitem{AnFangLin2022}
Dong An, Di~Fang, and Lin Lin.
\newblock Time-dependent hamiltonian simulation of highly oscillatory dynamics
  and superconvergence for schr{\"o}dinger equation.
\newblock {\em Quantum}, 6:690, 2022.

\bibitem{SharmaTran2024}
Kunal Sharma and Minh~C. Tran.
\newblock Hamiltonian simulation in the interaction picture using the magnus
  expansion, 2024.

\bibitem{BosseChildsEtAl2024}
Jan~Lukas Bosse, Andrew~M. Childs, Charles Derby, Filippo~Maria Gambetta,
  Ashley Montanaro, and Raul~A. Santos.
\newblock Efficient and practical hamiltonian simulation from time-dependent
  product formulas, 2024.

\bibitem{CasaresZiniArrazola2024}
Pablo Antonio~Moreno Casares, Modjtaba~Shokrian Zini, and Juan~Miguel Arrazola.
\newblock Quantum simulation of time-dependent hamiltonians via commutator-free
  quasi-magnus operators, 2024.

\bibitem{BlanesCasasOteoRos1998}
S~Blanes, F~Casas, J~A Oteo, and J~Ros.
\newblock Magnus and fer expansions for matrix differential equations: the
  convergence problem.
\newblock {\em Journal of Physics A: Mathematical and General}, 31(1):259, jan
  1998.

\bibitem{IserlesNorsett1999}
A.~Iserles and S.~P. Nørsett.
\newblock On the solution of linear differential equations in lie groups.
\newblock {\em Phil. Trans. R. Soc. A.}, 357:983--1019, 1999.

\bibitem{BlanesCasasRos2000}
S~Blanes, F~Casas, and J~Ros.
\newblock Improved high order integrators based on the magnus expansion.
\newblock {\em BIT Numerical Mathematics}, 40(3):434--450, 2000.

\bibitem{IserlesNorsettRasmussen2001}
A.~Iserles, S.P. Nørsett, and A.F. Rasmussen.
\newblock Time symmetry and high-order magnus methods.
\newblock {\em Applied Numerical Mathematics}, 39(3):379--401, 2001.
\newblock Themes in Geometric Integration.

\bibitem{HochbruckLubich2003}
Marlis Hochbruck and Christian Lubich.
\newblock On {M}agnus integrators for time-dependent {S}chr\"{o}dinger
  equations.
\newblock {\em SIAM J. Numer. Anal.}, 41(3):945--963, 2003.

\bibitem{Iserles2009}
A.~Iserles.
\newblock {\em A first course in the numerical analysis of differential
  equations}.
\newblock Number~44. Cambridge Univ. Pr., 2009.

\bibitem{BlanesCasasOteoRos2009}
S.~Blanes, F.~Casas, J.~A. Oteo, and J.~Ros.
\newblock The {M}agnus expansion and some of its applications.
\newblock {\em Phys. Rep.}, 470(5-6):151--238, 2009.

\bibitem{BlanesCasasOteoEtAl2010}
S~Blanes, F~Casas, J~A Oteo, and J~Ros.
\newblock {A pedagogical approach to the Magnus expansion}.
\newblock {\em Eur. J. Phys.}, 31(4):907--918, 2010.

\bibitem{IserlesKropielnickaSingh2017}
Arieh Iserles, Karolina Kropielnicka, and Pranav Singh.
\newblock Commutator-free magnus-lanczos methods for the linear schr{\"o}dinger
  equation, 2017.

\bibitem{BlanesCasasThalhammer2017}
Sergio Blanes, Fernando Casas, and Mechthild Thalhammer.
\newblock High-order commutator-free quasi-magnus exponential integrators for
  non-autonomous linear evolution equations.
\newblock {\em Computer Physics Communications}, 220:243--262, 2017.

\bibitem{IserlesKropielnickaSingh2018}
Arieh Iserles, Karolina Kropielnicka, and Pranav Singh.
\newblock Magnus--lanczos methods with simplified commutators for the
  schrödinger equation with a time-dependent potential.
\newblock {\em SIAM Journal on Numerical Analysis}, 56(3):1547--1569, 2018.

\bibitem{ChenForoozandehBuddSingh2023}
Guannan Chen, Mohammadali Foroozandeh, Chris Budd, and Pranav Singh.
\newblock Quantum simulation of highly-oscillatory many-body hamiltonians for
  near-term devices, 2023.

\bibitem{TranChuSuEtAl2020}
Minh~C Tran, Su-Kuan Chu, Yuan Su, Andrew~M Childs, and Alexey~V Gorshkov.
\newblock Destructive error interference in product-formula lattice simulation.
\newblock {\em Phys. Rev. Lett.}, 124(22):220502, 2020.

\bibitem{Layden2022}
David Layden.
\newblock First-order trotter error from a second-order perspective.
\newblock {\em Phys. Rev. Lett.}, 128:210501, May 2022.

\bibitem{PoulinQarrySommaVerstraete2011}
David Poulin, Angie Qarry, Rolando Somma, and Frank Verstraete.
\newblock Quantum simulation of time-dependent hamiltonians<? format?> and the
  convenient illusion of hilbert space.
\newblock {\em Physical review letters}, 106(17):170501, 2011.

\bibitem{Su2021}
Yuan Su.
\newblock Fast-{F}orwardable {Q}uantum {E}volution and {W}here to {F}ind
  {T}hem.
\newblock {\em {Quantum Views}}, 5:62, November 2021.

\bibitem{FangLinTong2023}
Di~Fang, Lin Lin, and Yu~Tong.
\newblock Time-marching based quantum solvers for time-dependent linear
  differential equations.
\newblock {\em {Quantum}}, 7:955, March 2023.

\bibitem{BerryCosta2022}
Dominic~W. Berry and Pedro C.~S. Costa.
\newblock Quantum algorithm for time-dependent differential equations using
  dyson series, 2022.

\bibitem{BurdenNA}
R.~L. Burden, J.~D. Faires, and A.~C. Reynolds.
\newblock {\em Numerical analysis}.
\newblock Brooks Cole, 2000.

\bibitem{SofferWu2020}
Avy Soffer and Xiaoxu Wu.
\newblock $l^p$ boundedness of the scattering wave operators of schroedinger
  dynamics with time-dependent potentials and applications, 2020.

\bibitem{Schlag2018}
W.~Schlag.
\newblock Intertwining wave operators, fourier restriction, and wiener
  theorems, 2018.

\bibitem{zworski2022semiclassical}
Maciej Zworski.
\newblock {\em Semiclassical Analysis}.
\newblock American Mathematical Society, 2012.

\bibitem{ReedSimon1975}
M.~Reed and B.~Simon.
\newblock {\em II: Fourier Analysis, Self-Adjointness}.
\newblock Number v. 2 in II: Fourier Analysis, Self-Adjointness. Elsevier
  Science, 1975.

\bibitem{GilyenSuLowEtAl2019}
Andr{\'a}s Gily{\'e}n, Yuan Su, Guang~Hao Low, and Nathan Wiebe.
\newblock Quantum singular value transformation and beyond: exponential
  improvements for quantum matrix arithmetics.
\newblock In {\em Proceedings of the 51st Annual ACM SIGACT Symposium on Theory
  of Computing}, pages 193--204, 2019.

\bibitem{sunderhauf2024block}
Christoph S{\"u}nderhauf, Earl Campbell, and Joan Camps.
\newblock Block-encoding structured matrices for data input in quantum
  computing.
\newblock {\em Quantum}, 8:1226, 2024.

\bibitem{camps2024explicit}
Daan Camps, Lin Lin, Roel Van~Beeumen, and Chao Yang.
\newblock Explicit quantum circuits for block encodings of certain sparse
  matrices.
\newblock {\em SIAM Journal on Matrix Analysis and Applications},
  45(1):801--827, 2024.

\bibitem{li2023efficient}
Haoya Li, Hongkang Ni, and Lexing Ying.
\newblock On efficient quantum block encoding of pseudo-differential operators.
\newblock {\em Quantum}, 7:1031, 2023.

\bibitem{babbush2018encoding}
Ryan Babbush, Craig Gidney, Dominic~W Berry, Nathan Wiebe, Jarrod McClean,
  Alexandru Paler, Austin Fowler, and Hartmut Neven.
\newblock Encoding electronic spectra in quantum circuits with linear t
  complexity.
\newblock {\em Physical Review X}, 8(4):041015, 2018.

\bibitem{Wan2021exponentially}
Kianna Wan.
\newblock Exponentially faster implementations of select (h) for fermionic
  hamiltonians.
\newblock {\em Quantum}, 5:380, 2021.

\bibitem{liu2024efficient}
Diyi Liu, Weijie Du, Lin Lin, James~P Vary, and Chao Yang.
\newblock An efficient quantum circuit for block encoding a pairing
  hamiltonian.
\newblock {\em arXiv preprint arXiv:2402.11205}, 2024.

\bibitem{du2024hamiltonian}
Weijie Du and James~P Vary.
\newblock Hamiltonian input model and spectroscopy on quantum computers.
\newblock {\em arXiv preprint arXiv:2402.08969}, 2024.

\bibitem{ChildsWiebe2012}
Andrew~M. Childs and Nathan Wiebe.
\newblock Hamiltonian simulation using linear combinations of unitary
  operations.
\newblock {\em Quantum Information and Computation}, 12, Nov 2012.

\bibitem{BerryChildsKothari2015}
D.~W. Berry, A.~M. Childs, and R.~Kothari.
\newblock Hamiltonian simulation with nearly optimal dependence on all
  parameters.
\newblock {\em Proceedings of the 56th IEEE Symposium on Foundations of
  Computer Science}, pages 792--809, 2015.

\bibitem{sanders2019black}
Yuval~R Sanders, Guang~Hao Low, Artur Scherer, and Dominic~W Berry.
\newblock Black-box quantum state preparation without arithmetic.
\newblock {\em Physical review letters}, 122(2):020502, 2019.

\bibitem{Oliveira2007QuantumBit}
David Oliveira and Rubens Ramos.
\newblock Quantum bit string comparator: Circuits and applications.
\newblock {\em Quantum Computers and Computing}, 7, 01 2007.

\bibitem{Martinez2002}
Andr{\'e} Martinez.
\newblock {\em {An introduction to semiclassical and microlocal analysis}},
  volume 994.
\newblock Springer, 2002.

\end{thebibliography}
\end{spacing}
\end{document}